**Thermodynamic modeling of binaries in Cr-Fe-Mo-Nb-Ni supported by first-principles calculations**

Hui Sun [a], Shun-Li Shang [a,*], Shuang Lin [a], Jingjing Li [b], Allison M. Beese [a, c], Zi-Kui Liu [a]

[a] Department of Materials Science and Engineering, Pennsylvania State University, University Park, PA 16802, USA

[b] Department of Industrial and Manufacturing Engineering, Pennsylvania State University, University Park, PA 16802, USA

[c] Department of Mechanical Engineering, Pennsylvania State University, University Park, PA 16802, USA

* Email: sus26@psu.edu

**Abstract:**

Thermodynamic descriptions of all binaries within the Cr-Fe-Mo-Nb-Ni system have been complied and, where necessary, remodeled. Notably, the Cr-Fe and Fe-Mo systems have been remodeled using comprehensive sublattice models for the topologically close-packed (TCP) phases of Laves_C14, $\sigma$, and $\mu$ according to their Wyckoff positions. These refinements are supported by first-principles calculations based on density functional theory (DFT), in conjunction with available experimental data in the literature. The resulting models offer improved accuracy in describing the TCP phases. For instance, the predicted site occupancies of $\sigma$ in Cr-Fe show excellent agreement with experimental observations. The present work provides a robust foundation for CALPHAD modeling and the design of complex, multi-component materials, particularly those based on Fe-based and Ni-based alloys.

**Highlights**

- Refined thermodynamic models for TCP phases in all binaries in Cr-Fe-Mo-Nb-Ni.

- DFT-assisted CALPAHD modeling of TCP phases of Laves_C14, $\sigma$, and $\mu$.

- Complete sublattice models of TCP phases applied to Cr-Fe and Fe-Mo systems.

- Robust foundation for future CALPHAD modeling and materials design.

# 1. Introduction

Topologically close-packed (TCP) phases, also known as Frank-Kasper phases [1], represent a category of intermetallic compounds (IMCs) with complex crystalline structures and high coordination numbers (up to 16). For instance, Table 1 summarizes the key TCP phases with their Wyckoff positions [2–6], including Laves_C14 (termed with C14 in this work) with space group $P6_3/mmc$ (no. 194) and three Wyckoff positions (2a, 4f, and 6h), Laves_C15 (termed with C15 in this work) with space group $Fd\bar{3}m$ (no. 227) and two Wyckoff positions (8a and 16d), δ with space group $Pmmn$ (no. 59) and three Wyckoff positions (2a, 2b, and 4f), μ with space group $R\bar{3}m$ (no. 166) and five Wyckoff positions (3a, $6c_1$, $6c_2$, $6c_3$, and 18h), and σ with space group $P4_2/mnm$ (no. 136) and five Wyckoff positions (2b, 4f, $8i_1$, $8i_2$, and 8j). These TCP phases are frequently observed in metallic alloys such as Ni-based [7], Fe-based [8], and high-entropy alloys [9].

TCP phases are not only inherently brittle, posing a threat to mechanical properties of alloys, but they also weaken solid solution strengthening by draining refractory elements from the matrix (like FCC-based γ phase in Ni-based superalloys [7]). For example, the formation of C14 in the P92 steel reduces Charpy toughness by 75 % after 3000 hours of aging [10]. Likewise, the formation of 10 vol.% of δ phase in Inconel 625 leads to a 45% drop in fracture strain after stress relief heat treatment [11]. Given their critical role on alloy performance, a deeper understanding and improved predictive modeling of TCP phases are essential for advanced alloy design,

Additive manufacturing (AM) offers a promising route for producing functionally graded materials (FGMs) from such as Ni-based superalloys to Fe-based superalloys with the key elements like Cr, Fe, Mo, Nb, and Ni [12]. However, understanding and design AM processes are currently hindered by the lack of a reliable thermodynamic database that spans the full compositional range — particularly in intermediate regions. Existing commercial databases such as TCFE and TCNI developed by Thermo-Calc [13] are primarily valid at the Fe- and Ni-rich corners, respectively. Therefore, a comprehensive thermodynamic database covering the entire composition space is essential for accurate modeling and design of Fe- and Ni-based superalloys.

The present work focuses on thermodynamic modeling of TCP phases across all 10 binaries within the Cr-Fe-Mo-Nb-Ni system. As summarized in in Table 2, seven TCP phases are considered in the present work, including δ, σ, μ, D_NiMo, C14, C15, and R [14–18]. In the CALPHAD modeling approach, a complete sublattice model for each phase can be built based on its Wyckoff positions, leading to an accurate description of phase stability and other properties. However, previous modeling works employed simplified sublattice models [19], assuming identical behavior of elements across multiple Wyckoff positions. For example, Jacob et al. [14] modelled the σ phase using a simplified representation of $(Cr,Fe)_4(Cr,Fe)_{10}(Cr,Fe)_{16}$ rather than the complete sublattice model of $(Cr,Fe)_1(Cr,Fe)_2(Cr,Fe)_4(Cr,Fe)_4(Cr,Fe)_4$. While simplification reduces computational complexity, it compromises accuracy in predicting site occupancies. For instance, the simplified sublattice model of μ phase $((Nb,Ni)_1Nb_4(Nb,Ni)_2(Nb,Ni)_6)$ in the Nb-Ni system used by Chen et al. [20] fails to reproduce experimental site fraction data [21]. In addition, an arbitrary value of 5000 J/mol-atom was assigned as the enthalpy of formation ($\Delta_f H$) for several endmembers in the previous modeling work [14]. Since ternary and higher-order systems are built upon binary descriptions, inaccuracies at the binary level propagate and undermine the fidelity of multicomponent modeling. Therefore, adopting complete sublattice models is essential for accurately capturing the thermodynamics of TCP phases in both binary and higher-order systems.

In the present work, a comprehensive review is conducted for all available binary modeling works within the Cr-Fe-Mo-Nb-Ni system. Both the Cr-Fe and Fe-Mo systems are specifically remodeled with complete sublattice models for the TCP phases, addressing the limitations of previously simplified models [14,15]. To support the remodeling, the $\Delta_f H$ values for all relevant endmembers are predicted by density functional theory (DFT) based first-principles calculations, providing reliable input data for the present CALPHAD modeling.

## 2. Overview of CALPHAD modeling of 10 binaries

### 2.1. Cr-Fe

The Cr-Fe system has four stable (including metastable, similarly hereinafter) phases, including liquid, BCC_A2, FCC_A1, and the TCP phase of σ. At least 13 CALPHAD modeling studies have been conducted on this system [14,22–33], with the most recent by Jacob et al. in 2018 [14]. They used a simplified three-sublattice model for the σ phase ((Cr, Fe)$_{10}$(Cr, Fe)$_4$(Cr, Fe)$_{16}$), which fails to reproduce the experimentally measured site fractions of Fe in the five Wyckoff positions at 973 K after annealing for up to 6650 h, as reported by Cieslak et al. [34]. While the modeling work by Xiong et al. [25] provided a better description of the Curie temperature, it relied on a revised lattice stability value for pure Fe, differing from the widely accepted values provided by the Scientific Group Thermodata Europe (SGTE) [35]. In the present work, the Cr-Fe system is hence remodeled using SGTE lattice stabilities and a complete five-sublattice model for σ, as detailed in Table 2. The remodeled Cr-Fe system not only improves the accuracy of site fraction predictions for the σ phase but also ensures compatibility for integration into the broader Cr-Fe-Mo-Nb-Ni thermodynamic database.

## 2.2. Cr-Mo

At least seven CALPHAD modeling studies have been conducted for the Cr-Mo system, which contains only two stable phases of liquid and BCC_A2 [36–42]. The most recent of these was performed by Jindal et al. in 2013 [42], employing the cluster expansion and cluster variation method (CE-CVM) to model BCC_A2 using the octahedron–rhombohedron–cube (ORC) approximation, it deviates from the widely adopted Redlich–Kister polynomial [43], which is more compatible with multicomponent CALPHAD databases. To ensure consistency and compatibility within the Cr-Fe-Mo-Nb-Ni system, the present work adopts the thermodynamic description by Frisk et al. [39] for the Cr-Mo system; see the calculated phase diagram in Fig. S1.

## 2.3. Cr-Nb

The Cr-Nb system contains three stable phases of liquid, BCC_A2, and Laves_C15, described by several CALPHAD modeling studies [16,44–49]. The most recent, conducted by Jiang et al. in 2018 [44] employed a segmented model that uses a weighted linear combination of Einstein functions to approximate heat capacity and describe the Debye model. This approach was applied to pure Cr and Nb to accurately capture their 0 K thermodynamic properties, as well as those of the stoichiometric $Cr_2Nb$ phase. However, as mentioned in Section 2.1, the present work adopts lattice stabilities from the SGTE for consistency across all binary systems. Since Jiang et al.'s model deviates from this convention, the second most recent model developed by Peng et al. in 2016 [16] is selected for the present work; see the calculated phase diagram in Fig. S2.

## 2.4. Cr-Ni

The Cr-Ni system has four stable phases of liquid, BCC_A2, FCC_A1, and $CrNi_2$ (Laves_C14). Earlier CALPHAD modeling efforts prior to 2013 [50–53] did not account for the low temperature Laves_C14 phase. Subsequent studies by Turchi et al. [54], Chan et al. [55], and Tang et al. [56] treated Laves_C14 as a stoichiometric compound, limiting its compositional flexibility. Hao et al. [57] introduced a two-sublattice model $((Cr, Ni)_1(Cr, Ni)_2)$ for Laves_C14, and adopted revised Gibbs energy functions for pure elements of Cr and Ni to improve the description of magnetic properties. However, these functions are incompatible with the widely accepted SGTE database, particularly regarding magnetic contributions. To maintain consistency with SGTE and ensure compatibility with multicomponent modeling, the present work adopts the thermodynamic description by Tang et al. [56], while remodeling of the Laves_C14 phase using a complete three-sublattice model for improved accuracy.

## 2.5. Fe-Mo

The Fe-Mo system comprises three solution phases of liquid, BCC_A2, and FCC_A1, and four TCP phases of R, σ, μ, and Laves_C14. Several CALPHAD modeling studies have been conducted for this system [15,58–62] using the simplified sublattice models for the TCP phases (σ, μ, and Laves_C14). For example, Rajkumar et al. in 2014 [15] used a two-sublattice model for σ $((Fe, Mo)_{10}(Fe, Mo)_{20})$, a four-sublattice model for μ $((Fe, Mo)_6(Fe, Mo)_2((Mo)_4(Fe, Mo)_1))$, and a two-sublattice model for Laves_C14 $((Fe, Mo)_1(Fe, Mo)_2)$. In contrast, the present work adopts complete models for all TCP phases based on their Wyckoff positions as detailed in Table 1. Specifically, a five-sublattice model is used for σ ((Fe, Mo)$_4$(Fe, Mo)$_4$(Fe, Mo)$_2$(Fe, Mo)$_4$((Fe, Mo)$_1$), a five-sublattice model for μ ((Fe, Mo)$_2$(Fe, Mo)$_2$(Fe, Mo)$_2$(Fe, Mo)$_6$((Fe, Mo)$_1$), and a three-sublattice model for Laves_C14 ((Fe, Mo)$_1$(Fe, Mo)$_1$(Fe, Mo)$_2$).

## 2.6. Fe-Nb

The Fe-Nb system includes two stable TCP phases (Laves_C14 and μ) as well as three stable solution phases (BCC_A2, FCC_A1, and liquid). Several CALPHAD modeling studies have been conducted for this system [19,63–71]. Here, we adopt the modeling work by Sun et al. [72] due to the complete sublattice models used for these two TCP phases based on their Wyckoff positions, i.e., the three-sublattice model for C14 (Fe,Nb)$_1$(Fe,Nb)$_2$(Fe,Nb)$_3$ and the five-sublattice model for μ (Fe,Nb)$_1$(Fe,Nb)$_2$-(Fe,Nb)$_2$(Fe,Nb)$_2$(Fe,Nb)$_6$.

## 2.7. Fe-Ni

The Fe-Ni system contains five stable phases, including liquid, FCC_A1, BCC_A2, FCC_L1$_2$, and FCC_L1$_0$. It has been modelled several times [73–82], and we adopt the most recent modeling work by Sun et al. in 2024 [72], who made minor adjustments to align the magnetic properties — specifically the Curie temperature (T$_C$) and the average magnetic moment ($\bar{\beta}$) from Ohnuma et al. [82] — of pure Ni with the values recommended by SGTE, ensuring compatibility with other binary systems in the database.

## 2.8. Mo-Nb

The Mo-Nb system consists of only two stable phases of liquid and BCC_A2. The present work adopts the latest modeling work by Yen et al. in 2020 [83], which demonstrates good agreement with both experimental data and DFT-based first-principles calculations..

## 2.9. Mo-Ni

The Mo-Ni system contains six stable phases of liquid, BCC_A2, FCC_A1, D_NiMo, MoNi$_4$, and δ. Several CALPHAD modeling studies have been conducted for this system [17,84–88]. Among those, the modeling work by Lei et al. in 2022 [17] is adopted in the present work, as it provides a better match with the enthalpy of mixing ($\Delta H_{mix}$) data for both liquid and solution phases than those by Yaqoob et al. [85] and Zhou et al. [84]. Note that we made minor modifications for the modeling work by Lei et al. [17] by replacing the original two-sublattice model for the δ phase with a three-sublattice model, based on its Wyckoff positions.

## 2.10. Nb-Ni

The Nb-Ni system includes six stable phases, i.e., three solution phases (FCC_A1, BCC_A2, and liquid) and three compounds (μ, δ, and NbNi$_8$). Six CALPHAD modeling studies have been reported for this system [21,61,89–92], with the most recent by Sun et al. in 2022 [93]. This model is adopted in the present work due to its use of complete sublattice models for the two TCP phases based on their Wyckoff positions. Specifically, a five-sublattice model $(Nb,Ni)_1(Nb,Ni)_2(Nb,Ni)_2$ $(Nb,Ni)_2(Nb,Ni)_6$ is used for μ and a three-sublattice model $(Nb,Ni)_1(Nb,Ni)_1(Nb,Ni)_2$ for δ. These models offer improved accuracy in reproducing the phase diagram and thermochemical properties in the Nb-Ni system [93].

# 3. Literature review of Cr-Fe and Fe-Mo

## 3.1. Thermochemical data

### 3.1.1. Cr-Fe

Xiong et al. [94] conducted a comprehensive review of experimental data available available for the Cr-Fe system prior to 2010. Four key types of thermodynamic properties are relevant – heat capacity, $\Delta H_{mix}$, activity, and site fraction. Heat capacity in the Cr-Fe system is significantly influenced by magnetic effects [14,94]. Since the present work adopts the SGTE database for pure elements, the modeling work by Jacob et al. [14], which also uses SGTE unary data, is adopted to describe heat capacity. This model aligns well with experimental results reported in [25,95,96].

Five experimental datasets are available for the entropy of mixing ($\Delta H_{mix}$) of liquid near 1960 K by Batalin et al. [97], Iguchi et al. [98], Thiedemann et al. [99], Pavars et al. [100], and Shumikhin et al. [101]. These datasets show significant discrepancies — up to 5 kJ/mol-atom. However, the data from Thiedemann et al. [99] and Pavars et al. [100] exhibit consistent trend, with difference of less than 0.5 with kJ/mol-atom. Therefore, only these two datasets are considered in the present work for liquid. For the BCC_A2 phase, two $\Delta H_{mix}$ datasets from Dench et al. [102] and Malinsky et al. [103] are available and show good agreement; both are included in the present modeling work.

As mentioned by Xiong et al. [94], experimental data of activity for the Cr-Fe system are scattered across different temperature ranges and derived using various experimental methods. Due to the lack of a standardized approach for evaluating these measurements, activity data are excluded from the present CALPHAD modeling.

Two sets of site fraction data for the σ phase are available from Cieslak et al. [34] and Yakel et al. [6], which are in good agreement. Both datasets are incorporated into the present modeling work.

### 3.1.2. Fe-Mo

For the Fe-Mo system, two types of experimental thermodynamic data are available: $\Delta H_{mix}$ and activity. There are three datasets of $\Delta H_{mix}$ values for liquid, all indicating negative deviations from ideal mixing. Among them, experimental data from Shumikhin et al. [101] exhibit unusual trends and significant uncertainty: $\Delta H_{mix}$ decreases to -0.21 kJ/mol-atom at 3.9 at. % Mo, increases to -0.17 kJ/mol-atom at 5.9 at. % Mo, and then decreases again to -0.34 kJ/mol-atom at 8.8 at. % Mo. Other datasets show notable discrepancies as well, for example, Sudavtsova et al. [104] reported $\Delta H_{mix}$ of 5 kJ/mol-atom at 20 at. % Mo, while Iguchi et al. [98] reported values of 2-3 kJ/mol-atom at 20-30 at. % Mo, a difference of approximately 2 kJ/mol-atom. Furthermore, Sudavtsova et al. [104] also measured the $\Delta H_{mix}$ of liquid (0 to -10.60 kJ/mol-atom from 0 to 20.0 at. % Nb) in the Fe-Nb system. These results show a large deviation (4 kJ/mol-atom) compared with other measurements [98,104,105], raising concerns about consistency. Due to these discrepancies, the present work considers only the experimental $\Delta H_{mix}$ data of liquid by Iguchi et al. [98] for the Fe-Mo liquid.

There are two sets of activity data from Ichise et al. [106] and Ueshima et al. [107], which are in good agreement. Therefore, both datasets are incorporated into the present CALPHAD modeling of the Fe-Mo system.

### 3.2. Phase equilibrium data

### 3.2.1. Cr-Fe

Extensive experimental data are available for the Cr-Fe system, as reviewed by Xiong et al. [94]. Based on their assessment, liquidus data from Putman et al. [108], Hellawell and Hume-Rothery [109], and Shurmann and Brauckmann [110] are selected in the present work. There are numerous measurements for the γ loop (i.e., the FCC_A1 phase) in the temperature range from 1100 to 1700 K and 0 - 15 at. % Fe [111–121]. These data exhibit scatter of approximately 60 K in temperature and 2.0 at. % Fe in composition. This variability is attributed to factors such as the use of impure raw materials (e.g., Cr of less than 100% purity), experimental uncertainties in temperature control, differences in sample preparation, measurement techniques, and potential instrument calibration errors. Given the lack of a systematic method to evaluate the reliability of these datasets within such a narrow compositional range, all available experimental data were considered in the present CALPHAD modeling.

For the phase boundary between BCC_A2 and σ, seven experiments studies are available [26,122–127]. Among these, the data from Dubiel and Inden [122] are considered the most reliable due to the exceptionally long annealing times (4 to 11 years). The data that consistent with Dubiel and Inden [122] are also used in the present work, specifically those from Novy et al. [123], Chandra and Schwartz [124], Pomey and Bastien [126], and Cook and Jones [127]. In contrast, the data from Hertzman and Sundman [26] are excluded due to shorter annealing times (18 months), which likely led to incomplete phase equilibration. Similarly, the work from De Nys and Gielen [125], which reported σ phase at 973 K — approximately 200 K higher than other studies — is excluded, likely due to insufficient annealing.

For the BCC_A2 miscibility gap, there are several experimental data from 690 to 900 K and 0 - 16 at. % Fe [122,128–132]. These data show significant internal scatter (up to 100 K) and discrepancies between different studies (up to 200 K). Therefore, only the data from Dubiel and Inden [122] are considered in the present work due to its longest annealing time.

### 3.2.2. Fe-Mo

A total of 19 experimental datasets are available for the Fe-Mo system [106,119,133–149], covering three solution phases of liquid, BCC_A2, and FCC_A1, and four TCP phases of R, $\sigma$, $\mu$, and C14, which were comprehensively reviewed by Rajkumar et al. [15]. Key experimental observations include that the observed R phase region measured by Heijwegen et al. [148] (34.6 – 38.2 at. % Mo, 1527 – 1581 K), the $\sigma$ phase region measured by Heijwegen et al. [148] (53.0 – 56.1 at. % Mo, 1557 – 1581 K), and the solidus temperatures 2191 – 2753 K with 60.0 – 90.0 at. % Mo by Rajkumar et al. [15]. The $\mu$ phase region has experimental data at 38.4 – 43.7 at. % Mo, 1137 – 1643 K, by Heijwegen et al. [148]. The C14 was identified by Rajkumar et al. [15] at 1123-1173 K. Across these studies, compositional discrepancies are generally within 5 at.% Mo, indicating good consistency among the datasets. Therefore, all available experimental data are incorporated into the present CALPHAD modeling work.

## 4. Methodology

### 4.1. DFT-based first-principles calculations

Helmholtz energy for a configuration of study can be predicted as a function of temperature $T$ and volume $V$ under a given pressure $P$ through the DFT-based quasiharmonic approach (QHA) [150]:

$$F(V,T) = E_0(V) + F_{vib}(V,T) + F_{el}(V,T) \qquad \text{Eq. 1}$$

where $F$ is the Helmholtz energy. $E_0(V)$ represents the static energy as a function of volume at 0 K excluding zero-point vibrational energy, $F_{vib}(V,T)$ the contribution of lattice vibrations, and $F_{el}(V,T)$ the contribution of thermal electrons. The equilibrium volume at each temperature is determined by $-\frac{\partial F}{\partial V} = P$. Here, we use $P = 0$ GPa and hence Helmholtz energy equals Gibbs energy.

$E_0(V)$ for each configuration, i.e., each endmember, is fitted using the 4-parameter Birch-Murnaghan (BM4) equation of state [150] after DFT-calculated energy-volume (E-V) data points at seven volumes near equilibrium,

$$E_0(V) = k_1 + k_2 V^{-2/3} + k_3 V^{-4/3} + k_4 V^{-2} \qquad \text{Eq. 2}$$

where $k_1$, $k_2$, $k_3$, and $k_4$ are fitting parameters, providing a means to derive material properties such as equilibrium volume, bulk modulus and its pressure derivative [150].

The phonon density of states (pDOS) are employed to estimate vibrational contribution [151],

$$F_{\text{vib}}(T,V) = k_B T \int_0^{\infty} \ln\left[2 \sinh\frac{\hbar\omega}{2k_B T}\right] g(\omega)\, d\omega \qquad \text{Eq. 3}$$

where $g(\omega)$ is pDOS as a function of $V$ and frequency $\omega$. At the same time, the Mermin statistics is used to compute the thermal electronic contribution, expressed as $F_{el} = E_{el} - TS_{el}$, where $E_{el}$ and $S_{el}$ are the internal energy of thermal electrons and the bare electronic entropy, respectively [151].

In the present work, all DFT-based first-principles calculations were performed by the Vienna *ab initio* Simulation Package (VASP) [152]. The projector augmented wave (PAW) method [153] was chosen to describe the electron-ion interaction while the generalized gradient approximation (GGA) by Perdew, Burke, and Ernzerhof (PBE) [154] was employed to describe the exchange-correlation functionals. For structural relaxations, a plane-wave cutoff energy of 368 eV was employed, and for the final static calculations to obtain accurate E-V data, a cutoff energy of 520 eV was used. The convergence criterion for electronic self-consistency in both relaxations and static calculations was set at $6 \times 10^{-5}$ eV/atom. The *k*-point meshes of ($2 \times 2 \times 4$), ($5 \times 5 \times 1$), and ($9 \times 9 \times 5$) were employed for both relaxations and static calculations for the $\sigma$, $\mu$, and C14 phases as shown in Table 3, respectively. The reference states were set as BCC Cr, BCC Mo, and BCC Fe. The numbers of endmembers for $\sigma$, $\mu$, and C14 are 243, 243, and 27, respectively. Crystal structures were sourced from the Materials Project database [155], as shown in the Table 1. The spin-polarization was used for Fe- and Cr-containing configurations, with antiferromagnetic spin-polarization used for BCC Cr.

All DFT-calculated values of enthalpy of formation for endmembers of $\sigma$, $\mu$, and C14 are shown in the supplementary thermodynamic database (TDB) file.

## 4.2. CALPHAD modeling

There are seven phases considered in the present work for the Cr-Fe and Fe-Mo systems, i.e., liquid, BCC_A2, FCC_A1, R, C14, $\sigma$, and $\mu$. The Gibbs energy functions for the pure elements (Cr, Fe, and Mo) were sourced from the SGTE database [35]. Both solution and non-stoichiometric phases were modeled with the compound energy formalism (CEF) [156].

The Gibbs energy for solution phases, including liquid, BCC_A2, and FCC_A1, is expressed using the Redlich-Kister polynomial [43],

$$G_m^\alpha = x_{Cr} G_{Cr}^\alpha + x_{Fe} G_{Fe}^\alpha RT(x_{Cr} ln x_{Cr} + x_{Fe} ln x_{Fe}) + x_{Cr} x_{Fe} \sum_{k=0} {}^k L_{Cr,Fe} (x_{Cr} - x_{Fe})^k \qquad \text{Eq. 4}$$

where $x_{Cr}$ and $x_{Fe}$ are mole fractions of Cr and Fe in phase $\alpha$. $G_{Cr}^\alpha$ and $G_{Fe}^\alpha$ are molar Gibbs energies of Cr and Fe relative to their stable-element reference (SER) states at 1 bar and 298.15 K. $R$ is the gas constant. ${}^k L_{i,j}$ represents the k$^{th}$ binary interaction parameter between elements $i$ and $j$, modelled by $a + bT$ with $a$ and $b$ being modeling parameters. Note that Eq. 4 uses the Cr-Fe system as an example, which is the same for the Fe-Mo system.

The R, C14, σ, and μ are non-stoichiometric TCP phases, as discussed in Section 3.2. The C14, σ, and μ were modeled with the sublattice models (CEF) based on their Wyckoff positions as shown in Table 1. While the R phase was modelled using a simplified sublattice model due to its complexity with 11 Wyckoff positions. The Gibbs energy per mole of formula unit for these TCP phases is given by,

$$G_{mf} = {}^0 G_{mf} + RT \sum_t a^t \sum_i y_i^t ln y_i^t + {}^E G_{mf} \qquad \text{Eq. 5}$$

where ${}^0 G_{mf}$ represents the Gibbs energy contribution from all endmembers, calculated by ${}^0 G_{mf} = \sum_{em}(\prod_t y_i^t \, {}^0 G_{em})$, where $y_i^t$ is the site fraction of component $i$ in sublattice $t$, ${}^0 G_{em}$ is the Gibbs energy of the corresponding endmember, and $a^t$ is the number of the sublattice sites. $\sum_i y_i^t ln y_i^t$ represents the ideal mixing entropy in each sublattice. ${}^E G_{mf}$ is the excess Gibbs energy containing two types of contributions: firstly, only one component in the non-mixing sublattices, while mixing of two or more components in one sublattice; secondly, the mixing of two or more components simultaneously in more than one sublattice. Only the first type is considered in this study since the second part is typically employed to describe short-range ordering [157]. The expression for the first type of interaction is:

$$^{E}G_{mf} = \sum_{t} \prod_{s \neq t} y_l^s \sum_{i>j} \sum_{j} y_i^t \, y_j^t \, L_{i,j:l}^t \qquad \text{Eq. 6}$$

where $L_{i,j:l}^t$ represents the interaction parameter between components $i$ and $j$ in sublattice $t$, while the other sublattices ($s$) featuring only single component in each sublattice.

PyCALPHAD modeling, fixed, modelled data.

## 5. Results and discussion

### 5.1. Cr-Fe

Fig. 1 shows the predicted site fractions of Fe in the σ phase of the Cr-Fe system at 973 K, based on the present CALPHAD modeling, superimposed with experimental data from Cieslak et al. [34]. The σ phase contains 5 Wyckoff sites, as listed in Table 1. Fig. 1 indicates that the present predictions show excellent agreement with measurements across all five sites. For example, the site fraction of Fe in site 2a ranges from 0.901-0.883, in site 4f from 0.253-0.236, in site $8i_1$ from 0.369–0.398, in site $8i_2$ from 0.890-0.907, and in site 8j from 0.280-0.350 within the composition range of 50.5 – 53.8 at. % Fe, aligning well with the measurements by Cieslak et al. [34] (site 2a ranges from 0.887-0.876, in site 4f from 0.249-0.296, in site $8i_1$ from 0.362–0.426, in site $8i_2$ from 0.900-0.900, and in site 8j from 0.285-0.357). Additionally, the mean absolute error (MAE) values from the present model are 0.022 for site 2a, 0.024 for site 4f, 0.011 for site $8i_1$, 0.011 for site $8i_2$, and 0.013 for site 8j, respectively, with respect to the measurements by Cieslak et al. [34]. As shown in Fig. S10, the present predictions of site fraction regarding Fe in σ at 923 K also match well with experimental data from Yakel et al. [6] with an overall MAE of 0.0288. In contrast, the other sublattice models for σ phase, such as the two-sublattice model by Xiong et al. [25] and the three-sublattice model by Jacob et al.

[14], are limited in their ability to resolve individual Wyckoff sites, as they combine multiple sites into fewer sublattices. This limits their predictive accuracy to only two or three sites.

Fig. 2 shows the enthalpy of formation for the BCC_A2 phase in the Cr-Fe system at 1529 K calculated using both the present CALPHAD modeling and that by Jacob et al. [14], alongside experimental data from Dench et al. [102] and Malinsky et al. [103]. Both modeling approaches capture the overall trend observed in the experimental data, particularly those from Dench et al. [102]. The enthalpy of formation of BCC increases to approximately 6 kJ/mol-atom at 50 at. % Fe then decreases to 2 kJ/mol-atom at 90.0 at. % Fe. The enthalpy of formation of BCC from the present model shows a deviation of 0.66 kJ/mol-atom from Dench et al. [102], and 0.56 kJ/mol-atom from Malinsky et al. [103]. In comparison, the model from Jacob et al. [14] exhibits a deviation of 0.49 kJ/mol-atom from Dench et al. [102], and only 0.07 kJ/mol-atom from Malinsky et al. [103]. The reported experimental uncertainty from Dench et al. [102] is approximately ±1.08 kJ/mol-atom. Overall, both CALPHAD modeling works demonstrate reasonable agreement with the experimental data, effectively capturing the compositional dependence of the enthalpy of formation in BCC_A2.

Fig. 3 shows the calculated $\Delta H_{mix}$ values of liquid at 1960 K for the Cr-Fe system from both the present CALPHAD modeling and that from Jacob et al. [14], in comparison with experimental data [97–101]. As discussed in Section 3.1, significant discrepancies exist among the experimental datasets, with difference of up to 5 kJ/mol-atom. Here, the data from Thiedemann et al. [99] and Pavars et al. [100] are considered in the present work since they agree well with each other. The present model predicts $\Delta H_{mix}$ values of liquid close 0 kJ/mol-atom from 0-100 at. % Fe, aligning well with the selected experimental data. The deviations from Thiedemann et al. [99] and Pavars et al. [100] are only 0.19 and 0.02 kJ/mol-atom, respectively. In contrast, the model by Jacob et al. [14] shows average deviations of 0.48 and 0.27 kJ/mol-atom from the same datasets.

Fig. 4 (a) displays the calculated Cr-Fe phase diagram the present CALPHAD modeling, while Fig. 4 (b) provides a zoomed-in view focusing on the 0-20 at. % Fe range and 800-1800 K. The phase diagram includes four phases of liquid, BCC_A2, FCC_A1, and $\sigma$. The invariant reactions are summarized in Table 3: Details of DFT-based first-principles calculations for each compound (endmember) or element, including total atoms in the unit cells, $k$-points meshes for structure relaxations and the final static calculations.

| Compounds | Atoms in the cells | $k$-points meshes |
|---|---|---|
| BCC-Cr | 2 | 31×31×31 |
| BCC-Fe | 2 | 31×31×31 |
| BCC-Mo | 2 | 31×31×31 |
| $\sigma$ | 30 | 2×2×4 |
| $\mu$ | 39 | 5×5×1 |
| C14 | 12 | 9×9×5 |

Table 4, including three congruent reactions and one eutectic reaction. As shown in Fig. 4 (b), experimental data of the $\gamma$ loop (FCC_A1 phase) between 1100 - 1700 K and 0-15 at. % Fe reveal a ~60 K discrepancy between measurements by Roe et al. [113] and Adcock et al. [158]. Therefore, the FCC_A1 to BCC_A2 phase transition predicted by the present model is 1111 K, while Jacob et al. [14] predicts 1126 K. Both values fall within the experimental range, with the present model aligning more closely with Adcock et al. [158] (1107 K), and Jacob et al. [14] aligning better with Oberhoffer and Esser [158] (1134 K). For the eutectic reaction ($\sigma \rightarrow$ BCC_A2 + BCC_A2', where BCC_A2' denotes the second BCC_A2 phase due to a miscibility gap), experimental temperatures range from 773 – 783 K [122,123]. The present model predicts

772 K, slightly closer to the experimental values than Jacob et al. [14], who predict 788 K. Both models also show excellent agreement with experimental data for the other two congruent reactions (liquid → BCC_A2 at 21 at. % Fe and 1783 K [158], and BCC_A2 → σ at 47 at. % Fe and 1093 K [127]). In both cases, the temperature differences are less than 1 K and the composition differences are under 1 at.% Fe.

## 5.2. Fe-Mo

Fig. 5 presents the predicted $\Delta H_{mix}$ values of liquid at 1848 K for the Fe-Mo system from both the present CALPHAD model and the model by Rajkumar et al. [15], in comparison with experimental data [98]. The present predictions match well with the experimental data by Iguchi et al. [98], yielding a MAE of 0.55 kJ/mol-atom, while the model by Rajkumar et al. [15] gives a similar MAE of 0.56 kJ/mol-atom. However, a notable discrepancy of approximately 1 kJ/mol-atom exists at 60 at. % Mo between two models. It is important to note that no experimental $\Delta H_{mix}$ data are available for the 30-100 at. % Mo range. Nevertheless, as shown in Fig. 7, the present model more accurately predicts the liquid–BCC_A2 phase boundaries in the 60–90 at.% Mo range, supporting its overall reliability.

Fig. 6 shows the predicted activity of Fe in the Fe-Mo system at 1823 K from both the present model and that of Rajkumar et al. [15], superimposed with experimental data from Ichise et al. [106] and Ueshima et al. [107]. Both models reproduce the experimental trend: Fe activity decreases from 1.0 at 0 at.% Mo, stabilizes around 0.61 between 35–48 at.% Mo, drops to 0.52 at 56 at.% Mo, and decreases further at 81 at.% Mo. The present model yields MAEs of 0.029 and 0.030 with respect to Ichise et al. [106] and Ueshima et al. [107], respectively, while Rajkumar et al.'s model shows slightly lower MAEs of 0.024 and 0.017. These results indicate that both models are consistent with experimental observations.

Fig. 7 (a) displays the calculated Fe-Mo phase diagram from the present CALPHAD modeling, while Fig. 7 (b) provides a zoomed-in view for 0-4 at. % Mo and 1000-1800 K. The system includes seven phases of liquid, BCC_A2, FCC_A1, R, σ, μ, and C14; and five peritectic reactions and eutectic reactions as summarized in Table 5. For the peritectic reaction (liquid → BCC_A2 + σ), both the present model predicts 43.9 at. % Mo in liquid, 75.2 at. % Mo in BCC_A2, and 53.0 at. % Mo in σ at 1888 K. Rajkumar et al. [15] predicts 40.6 at. % Mo in liquid, 73.4 at. % Mo in BCC_A2, and 54.0 at. % Mo in σ at 1881 K. These predictions align well with experimental data from Ueshima et al. [145] (39.4 at. % Mo in liquid, 75.2 at. % Mo in BCC_A2, and 54.2 at. % Mo in σ at 1883 K). The present model better captures the BCC_A2 composition, while Rajkumar et al.'s model aligns more closely with the liquid and σ compositions. For the peritectic reaction (liquid → σ + R), the present model predicts 30.4 at. % Mo in liquid, 44.7 at. % Mo in σ, and 39.5 at. % Mo in R at 1766 K. Rajkumar et al. [15] predicts 29.3 at. % Mo in liquid, 47.9 at. % Mo in σ, and 36.1 at. % Mo in R at 1760 K. These results are consistent with the reaction temperature reported by Gibson et al. [138] (1761 K). The present model better matches the measured compositions of liquid [138] and σ [140] (41.0 at. % Mo), while Rajkumar et al.' model [15] aligns better with the R phase composition (37.4 at.% Mo) from Sinha et al. [134]. Additional reactions include liquid + R → BCC_A2 at 1722 K [138], R + σ → μ at 1643 [138], σ → μ + BCC_A2 at 1507-1520 K [148], R → μ + BCC_A2 at 1473 K [148], and BCC_A2 + μ → C14 at 1123-1223 K [15,148]. Both models agree well with experimental temperatures for these reactions, with deviations under 3 K.

As shown in Fig. 7 (a and b), the experimental phase boundary for the R phase spans 34.6 – 38.2 at. % Mo and 1527 – 1581 K [148]. The present model predicts 34.4 – 36.4 at. % Mo, closely matching the data, while Rajkumar et al. [15] predicts 36.0 – 36.4 at. % Mo. Similarly, for the σ phase, experimental boundaries are 53.0–56.1 at.% Mo and 1557–1581 K [148], which align better with the present model (53.7–55.6 at.% Mo) than with Rajkumar et al. [15] (53.8–55.3 at.% Mo). Furthermore, the present model predicts solidus temperatures from 59.7 at.% Mo at 2191 K to 91.1 at.% Mo at 2753 K, more accurately reflecting experimental trends than Rajkumar et al. [15], who predict 55.6–89.8 at.% Mo over the same temperature range. This is also more consistent with the reported experimental range of 60.0-90.0 at.% Mo.

Overall, the present CALPHAD modeling demonstrates improved agreement with experimental data across $\Delta H_{mix}$, activity, phase boundaries, and solidus temperatures, offering a more reliable thermodynamic description of the Fe-Mo system compared to the model by Rajkumar et al. [15].

## 6. Conclusions

The present work significantly advances the thermodynamic modeling of the Cr-Fe-Mo-Nb-Ni system by refining the binary subsystems Cr-Fe and Fe-Mo, as well as by adjusting the models in Cr-Ni and Mo-Ni. This is achieved through the integration of density functional theory (DFT) calculations, experimental data, and comprehensive modeling of three topologically close-packed (TCP) phases of C14, σ, and μ using complete sublattice models. Furthermore, DFT calculations provide reliable thermodynamic properties at 0 K for all the endmembers, while experimental data from the literature serve to validate and refine the models. Together, these efforts establish a robust thermodynamic foundation for the Cr-Fe-Mo-Nb-Ni system, supporting both materials design and future CALPHAD modeling of higher-order multicomponent systems.

## Acknowledgments:

The authors thank Dr. Wei Xiong for sharing all experimental data in the Cr-Fe system. The authors acknowledge financial support by the Army Research Office (ARO) with Contract No. W911NF2320143, Office of Naval Research (ONR) under Contract No. N00014-21-1-2608, and National Science Foundation (NSF) via Award No. CMMI-2050069. First-principles calculations were performed partially on the Roar supercomputers at the Pennsylvania State University's Institute for Computational and Data Sciences (ICDS), partially on the resources of the National Energy Research Scientific Computing Center (NERSC), a DOE Office of Science User Facility supported under Contract No. DE-AC02-05CH11231 using the NERSC award BES-ERCAP0032760, and partially on the resources of Advanced Cyberinfrastructure Coordination Ecosystem: Services & Support (ACCESS) through allocation DMR1400063, which is supported by U.S. National Science Foundation Grants Nos. 2138259, 2138286, 2138307, 2137603, and 2138296.

## Tables:

Table 1. Crystallographic strucutre and Wyckoff positions of the key TCP phases of C14, C15, $\delta$, $\mu$, and $\sigma$ based on experiments.

| TCP Phases | Positions | x | y | z |
|---|---|---|---|---|
| C14 [a] | 2a | 0 | 0 | 0 |
| | 6h | 0.830 | 0.660 | 0.25 |
| | 4f | 0.667 | 0.333 | 0.933 |
| C15 [b] | 8a | 0.125 | 0.125 | 0.125 |
| | 16d | 0.5 | 0 | 0.5 |
| $\delta$ (delta) [c] | 2a | 0 | 0 | 0.318 |
| | 2b | 0 | 0.5 | 0.651 |
| | 4f | 0.75 | 0 | 0.841 |
| $\mu$ (mu) [d] | 3a | 0 | 0 | 0 |
| | 6c (1) | 0 | 0 | 0.167 |
| | 6c (2) | 0 | 0 | 0.346 |
| | 6c (3) | 0 | 0 | 0.448 |
| | 18h | 0.5 | 0.5 | 0.590 |
| $\sigma$ (sigma) [e] | 2b | 0 | 0 | 0 |
| | 4f | 0.101 | 0.899 | 0 |
| | 8i (1) | 0.037 | 0.631 | 0.5 |
| | 8i (2) | 0.066 | 0.739 | 0 |
| | 8j | 0.183 | 0.183 | 0.252 |

[a] Laves C14 with space group P6$_3$/mmc (no. 194) and prototype of MgZn$_2$ [2].

[b] Laves C15 with space group Fd$\bar{3}$m (no. 227) and prototype of Cu$_2$Mg [3].

[c] $\delta$ with space group Pmmn (no. 59), Strukturbericht designation D0$_a$, and prototype of $\beta$-Cu$_3$Ti [4].

[d] $\mu$ with space group R$\bar{3}$m (no. 166), Strukturbericht designation D8$_5$, and prototype of Fe$_7$W$_6$ [5].

[e] $\sigma$ with space group P4$_2$/mnm (no. 136), Strukturbericht designation D8$_b$, and prototype of $\sigma$-CrFe [6].

Table 2: Sublattice models used by the present work and those in the literature.

| Phases [a] | This work | Jacob et al. [14] | Rajkumar et al. [15] |
|---|---|---|---|
| BCC_A2 | $(Cr,Fe,Mo)_1(Va)_3$ | $(Cr,Fe)_1(Va)_3$ | $(Fe,Mo)_1(Va)_3$ |
| FCC_A1 | $(Cr,Fe,Mo)_1(Va)_3$ | $(Cr,Fe)_1(Va)_3$ | $(Fe,Mo)_1(Va)_3$ |
| C14 | $(Cr,Fe,Mo)_1(Cr,Fe,Mo)_2$ $(Cr,Fe,Mo)_3$ | | $(Fe,Mo)_1(Fe,Mo)_2$ |
| liquid | $(Cr,Fe,Mo)_1$ | $(Cr,Fe)_1$ | $(Fe,Mo)_1$ |
| μ | $(Cr,Fe,Mo)_1(Cr,Fe,Mo)_2$ $(Cr,Fe,Mo)_2(Cr,Fe,Mo)_2$ $(Cr,Fe,Mo)_6$ | | $(Fe,Mo)_6(Fe,Mo)_2$ $(Mo)_4(Fe,Mo)_1$ |
| σ | $(Cr,Fe,Mo)_1(Cr,Fe,Mo)_2$ $(Cr,Fe,Mo)_4(Cr,Fe,Mo)_4$ $(Cr,Fe,Mo)_4$ | $(Cr,Fe)_{10}(Cr,Fe)_4(Cr,Fe)_{16}$ | $(Fe,Mo)_{10}(Fe,Mo)_{20}$ |
| R | $(Cr,Fe)_{17}(Cr,Mo)_{14}(Cr,Fe,Mo)_{12}$ | | $(Fe)_{32}(Mo)_{18}(Fe,Mo)_3$ |

[a] The other three TCP phases and their models used herein are the same as those in the literature, including D_NiMo for the Ni-Mo system [17], δ for the Nb-Ni system [18], and C15 for the Cr-Nb system [16].

Table 3: Details of DFT-based first-principles calculations for each compound (endmember) or element, including total atoms in the unit cells, *k*-points meshes for structure relaxations and the final static calculations.

| Compounds | Atoms in the cells | *k*-points meshes |
|-----------|--------------------|--------------------|
| BCC-Cr | 2 | 31×31×31 |
| BCC-Fe | 2 | 31×31×31 |
| BCC-Mo | 2 | 31×31×31 |
| σ | 30 | 2×2×4 |
| μ | 39 | 5×5×1 |
| C14 | 12 | 9×9×5 |

Table 4: Predicted invariant reactions in the Cr-Fe system by CALPHAD modeling in comparison with available experiments in the literature (indicated by Expt.).

| Type | Compositions (at. % Fe) | | | | Temp. (K) | Source |
|------|------|------|------|------|-----------|--------|
| Congruent | Liquid | ↔ | BCC_A2 | | | |
| | 21.5 | | 21.5 | | 1782 | This study |
| | 21.0 | | 21.0 | | 1782 | Jacob et al. [14] |
| | 21.0 | | 21.0 | | 1783 | Expt. [158] |
| Congruent | FCC_A1 | ↔ | BCC_A2 | | | |
| | 7.5 | | 7.5 | | 1111 | This study |
| | 7.3 | | 7.3 | | 1126 | Jacob et al. [14] |
| | 7.0 | | 7.0 | | 1119 | Xiong et al. [25] |
| Congruent | BCC_A2 | ↔ | σ | | | |
| | 47.2 | | 47.2 | | 1095 | This study |
| | 46.0 | | 46.0 | | 1096 | Jacob et al. [14] |
| | 47.0 | | 47.0 | | 1093 | Expt. [127] |
| Eutectic | σ | ↔ | BCC_A2 | + | BCC_A2' | | |
| | 48.0 | | 16.4 | | 84.9 | 772 | This study |
| | 49.0 | | 15.0 | | 85.0 | 788 | Jacob et al. [14] |
| | 48.6±0.6 | | 14.3 | | 88.0 | 773-783 | Expt. [122] |
| | | | 14.0 | | 83.0±1.0 | 773 | Expt. [123] |

Table 5: Predicted invariant reactions in the Fe-Mo system by CALPHAD modeling in comparison with available experiments in the literature (indicated by Expt.).

| Type | Compositions (at. % Mo) | | | | | Temp. (K) | Source |
|---|---|---|---|---|---|---|---|
| Peritectic | Liquid | + | BCC_A2 | ↔ | σ | | |
| | 43.9 | | 75.2 | | 53.0 | 1888 | This study |
| | 40.6 | | 73.4 | | 54.0 | 1882 | Rajkumar et al. [15] |
| | 39.4 | | 75.2 | | 54.2 | 1883 | Expt. [145] |
| | | | | | | 1815 | Expt. [159] |
| | | | | | | 1815 | Expt. [160] |
| | | | | | | 1815 | Expt. [140] |
| Peritectic | Liquid | + | σ | ↔ | R | | |
| | 30.4 | | 44.7 | | 39.5 | 1766 | This study |
| | 29.3 | | 47.9 | | 36.1 | 1760 | Rajkumar et al. [15] |
| | | | | | 37.4 | | Expt. [134] |
| | 28.0 | | | | | 1761 | Expt. [138] |
| | | | | | 41.0 | | Expt. [140] |
| Peritectic | Liquid | + | R | ↔ | BCC_A2 | | |
| | 25.0 | | 37.7 | | 23.7 | 1723 | This study |
| | 23.7 | | 35.8 | | 23.2 | 1721 | Rajkumar et al. [15] |
| | | | | | | 1722 | Expt. [138] |
| Peritectic | R | + | σ | ↔ | μ | | |
| | 40.2 | | 50.9 | | 41.1 | 1646 | This study |
| | 36.5 | | 49.9 | | 41.1 | 1645 | Rajkumar et al. [15] |
| | | | | | | 1643 | Expt. [138] |
| Eutectic | σ | ↔ | μ | + | BCC_A2 | | |
| | 56.0 | | 43.3 | | 94.6 | 1513 | This study |
| | 55.5 | | 44.3 | | 94.2 | 1514 | Rajkumar et al. [15] |
| | | | | | | 1513 | Expt. [15] |
| | | | | | | 1507-1520 | Expt. [148] |
| Eutectic | R | ↔ | μ | + | BCC_A2 | | |
| | 33.5 | | 38.2 | | 12.5 | 1476 | This study |
| | 36.1 | | 39.2 | | 13.2 | 1473 | Rajkumar et al. [15] |
| | | | | | | 1473 | Expt. [148] |
| | | | | | | 1473 | Expt. [144] |
| Peritectic | BCC_A2 | + | μ | ↔ | C14 | | |

| | | | | |
|---|---|---|---|---|
| 4.6 | 39.9 | 33.3 | 1179 | This study |
| 5.7 | 41.7 | 33.7 | 1200 | Rajkumar et al. [15] |
| | | | 1123-1173 | Expt. [15] |
| | | | 1223 | Expt. [148] |

**Figures:**

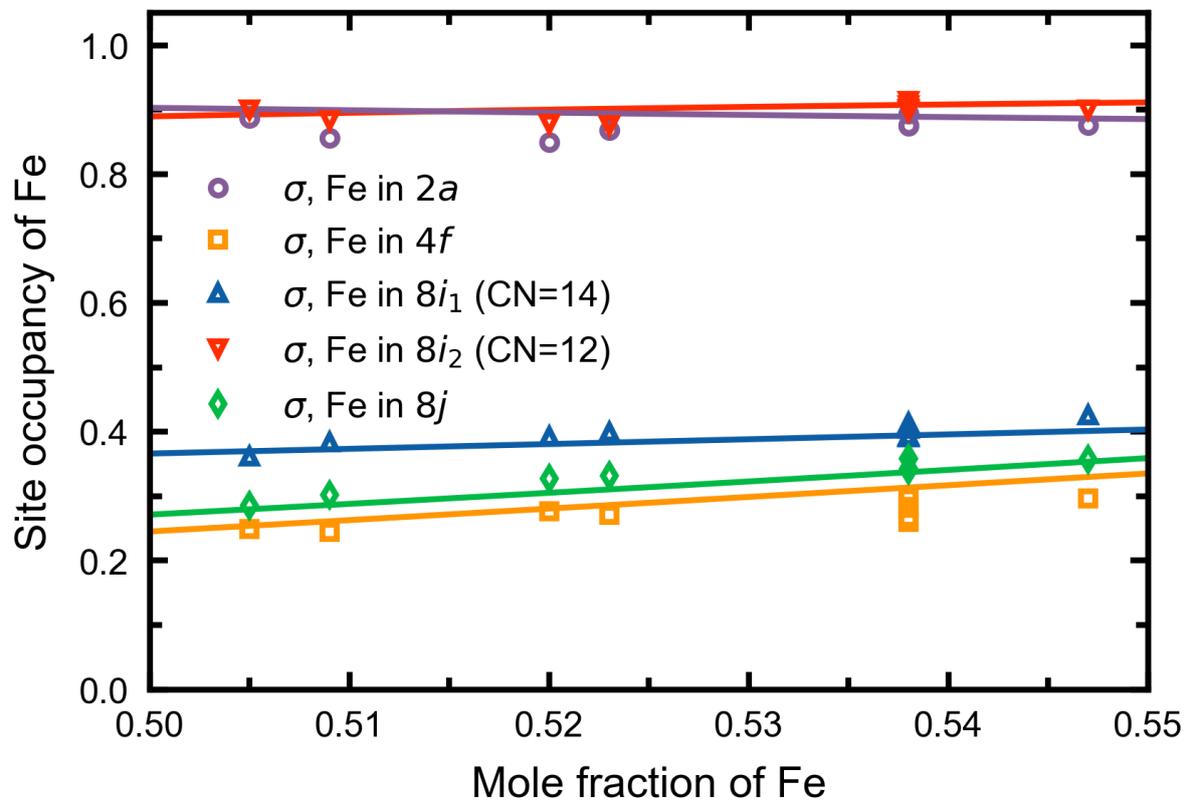

Fig. 1. Calculated site fractions of Fe in σ phase in the Cr-Fe system at 973 K by the present CALPHAD modeling, in comparison with experimental data from Cieslak et al. [34].

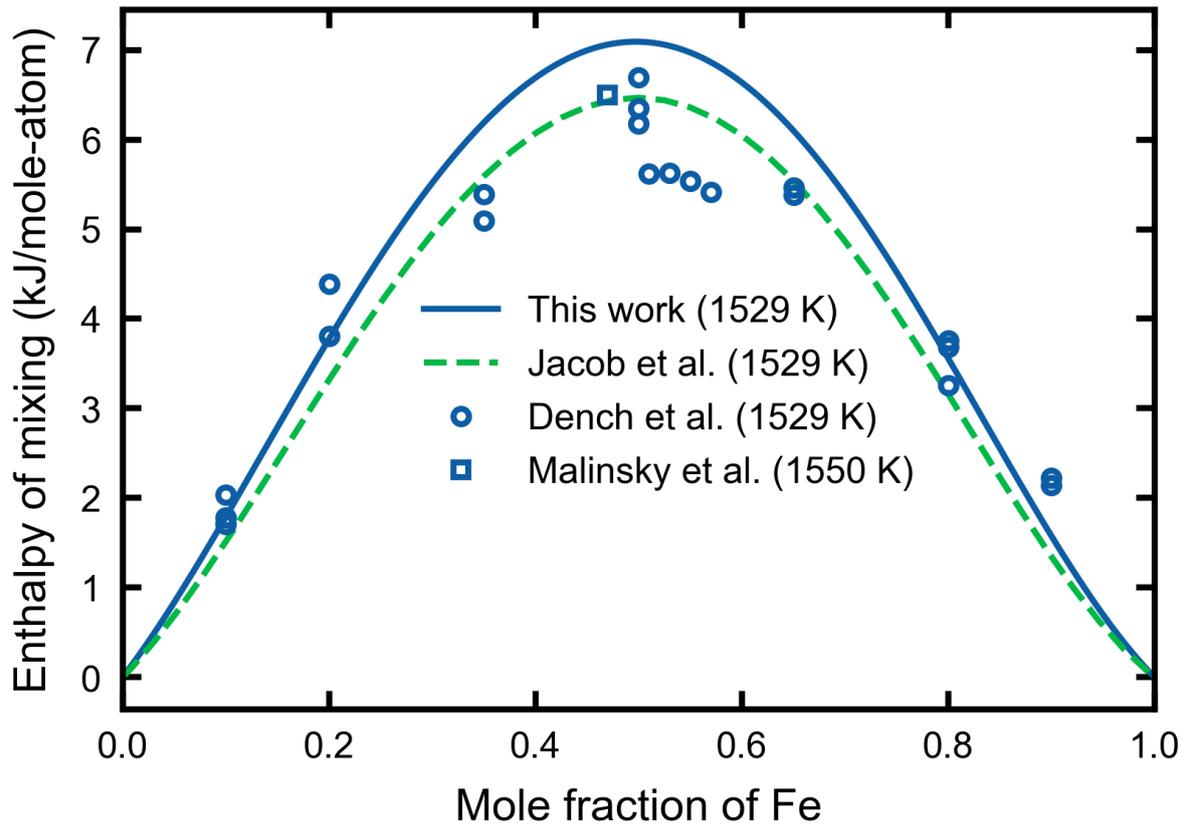

Fig. 2. Calculated enthalpy of formation for BCC_A2 in the Cr-Fe system at 1529 K from both the present CALPHAD modeling and the modeling by Jacob et al. [14], compared with experimental data from Dench et al. [102] and Malinsky et al. [103].

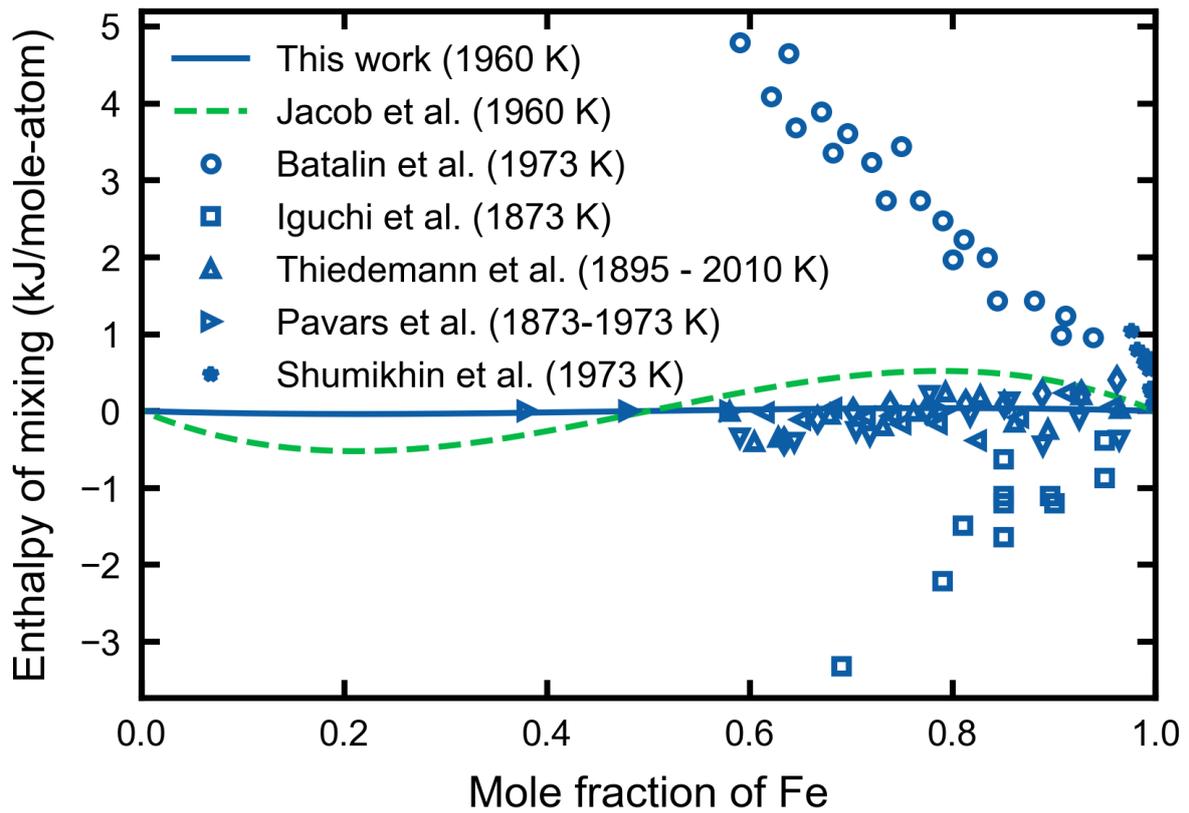

Fig. 3. Calculated enthalpy of mixing of liquid at 1960 K in the Cr-Fe system from both the present CALPHAD modeling and the modeling from Jacob et al. [14] with experimental data from Batalin et al. [97], Iguchi et al. [98], Thiedemann et al. [99], Pavars et al. [100], and Shumikhin et al. [101].

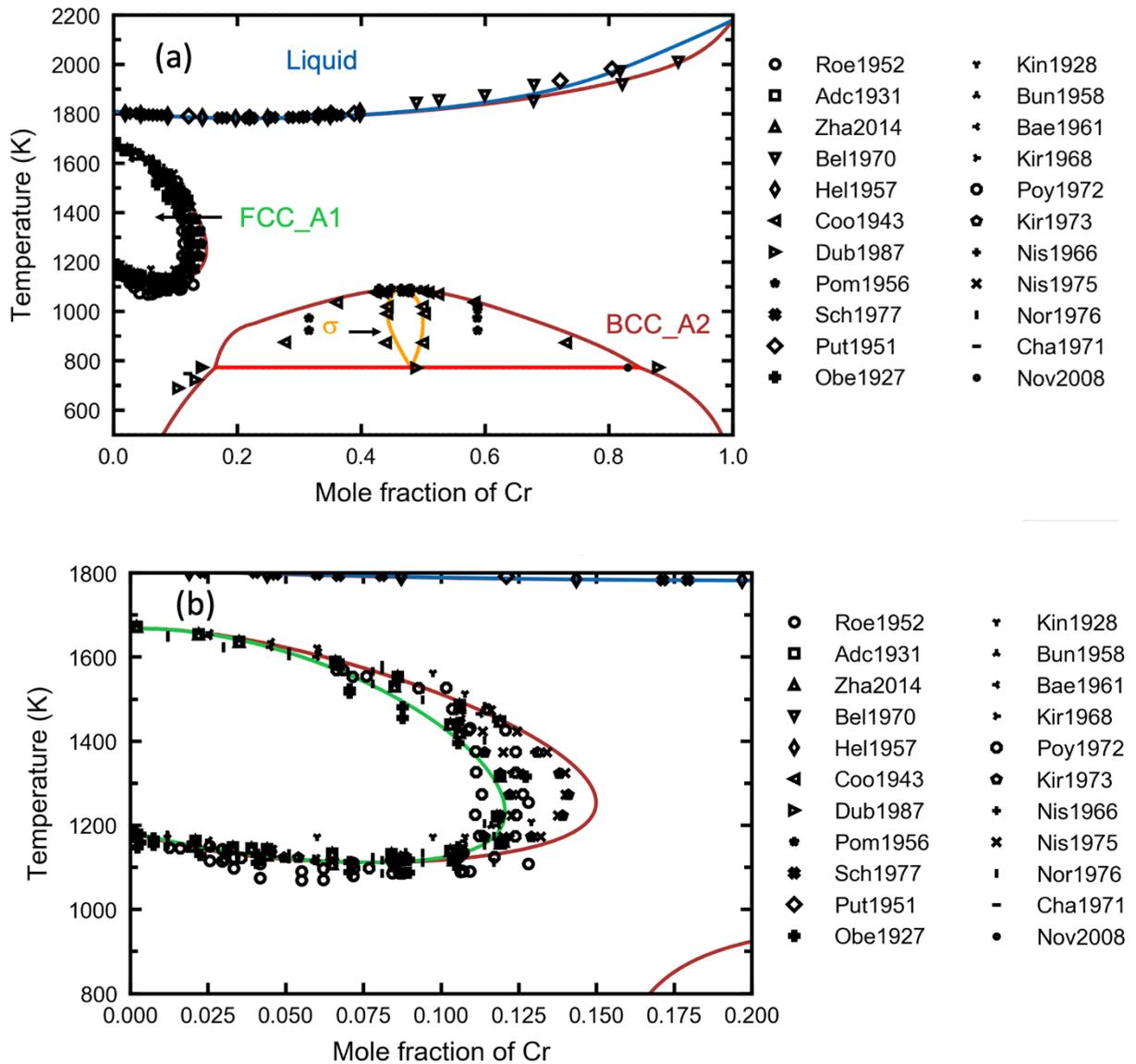

Fig. 4. (a) Calculated phase diagram of the Cr-Fe system based on the present CALPHAD modeling in comparison with experimental data (symbols, see details in [14,25]). (b) Zoomed-in phase diagram of the Cr-Fe system.

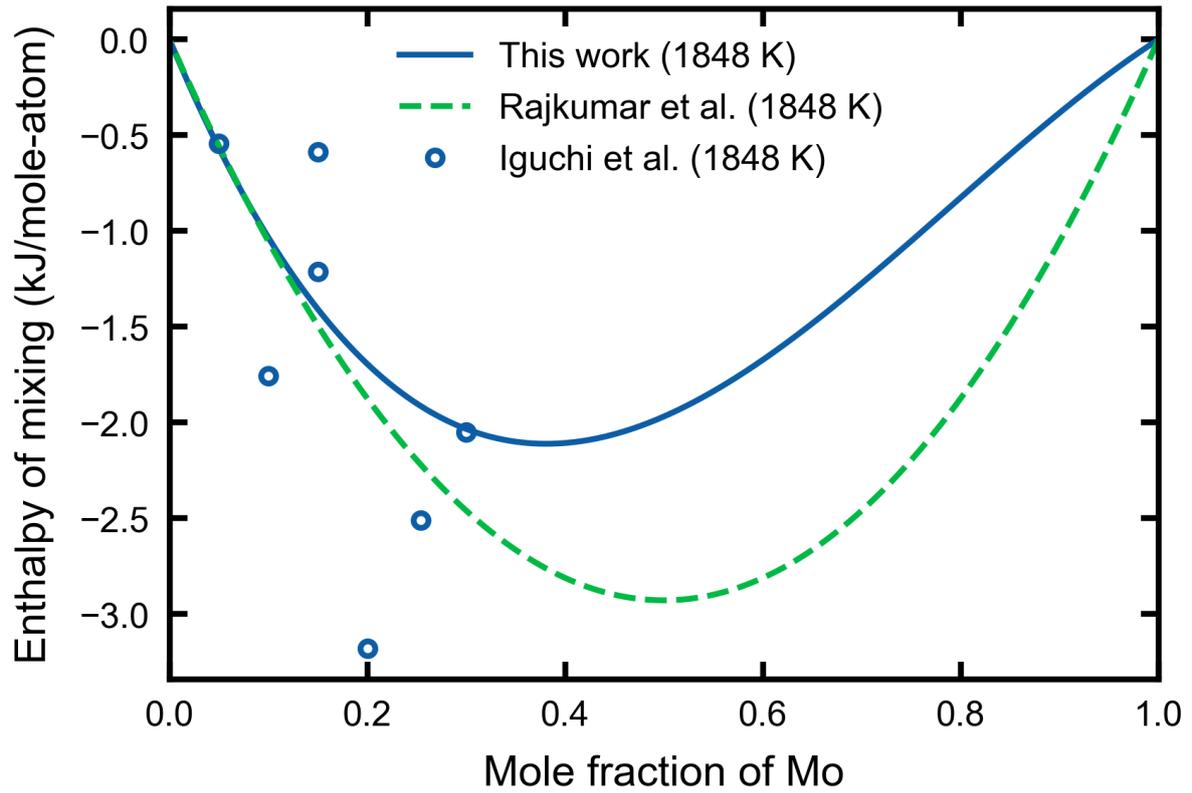

Fig. 5. Enthalpy of mixing of liquid at 1848 K for the Fe-Mo system calculated from both the present CALPHAD modeling and the modeling from Rajkumar et al. [15] with experimental data from Iguchi et al. [98].

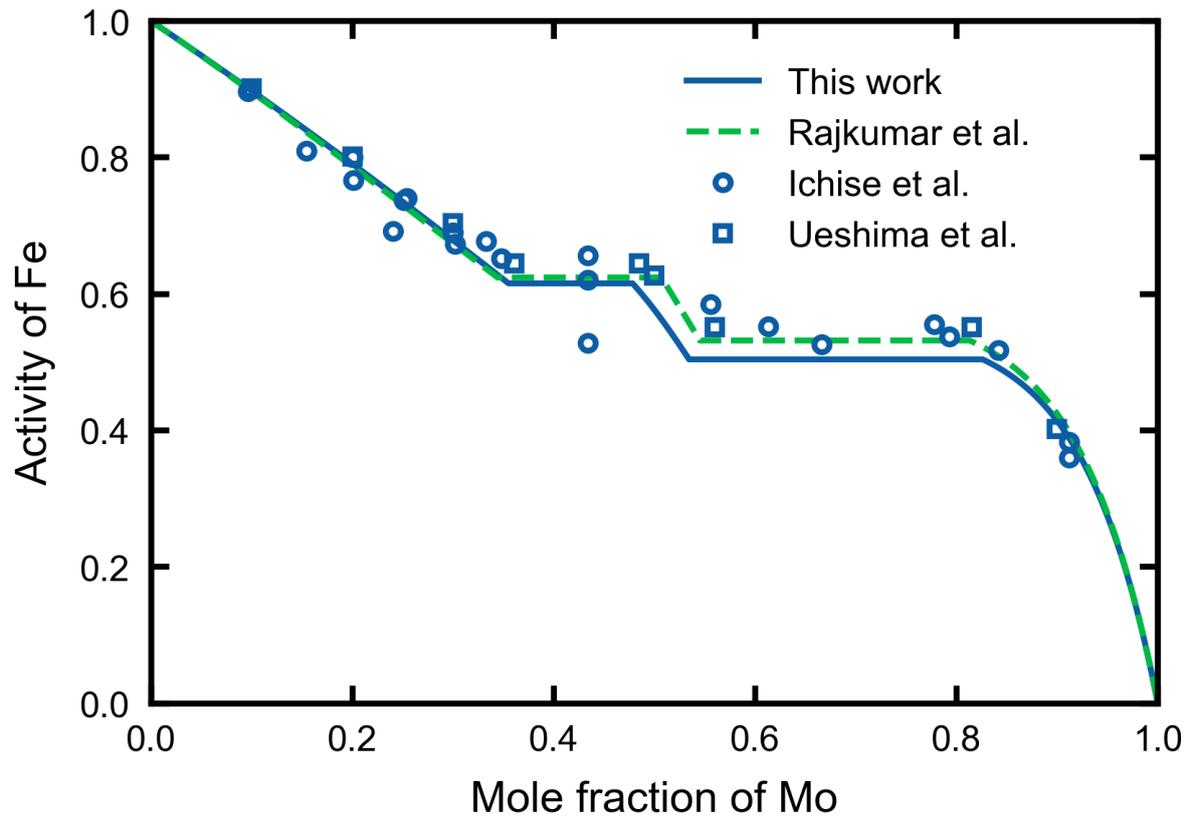

Fig. 6. Activity of Fe at 1823 K for the Fe-Mo system calculated from both the present CALPHAD modeling and the modeling from Rajkumar et al. [15] with experimental data from Ichise et al. [106] and Ueshima et al. [107].

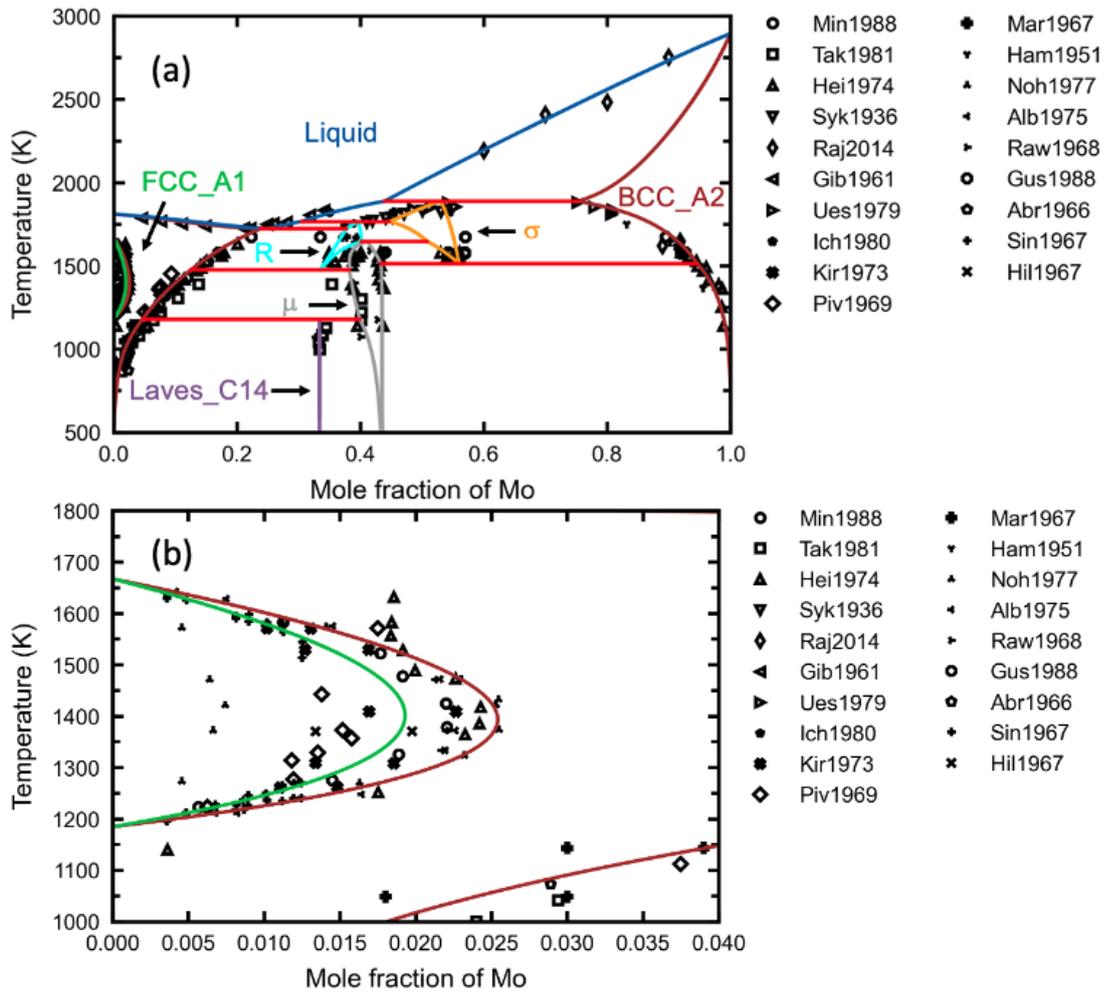

Fig. 7. (a) Calculated phase diagram of the Fe-Mo system based on the present CALPHAD modeling in comparison with experimental data (symbols, see details in [15]). (b) Zoomed-in phase diagram of the Fe-Mo system.

# 7. Supplemental Material

The supplemental material includes supplementary <mark>figures</mark> and one thermodynamic database (TDB) file as shown in the following pages.

**Fig. S1.** shows the predicted phonon density of states of the BCC-Nb phase (blue line), the FCC-Ni phase (green line), using DFT-based phonon calculations in comparison with experimental data [161].

**Fig. S2.** shows the presently predicted $\Delta H_{form}$ of all configurations of $\mu$-$Nb_7Ni_6$ (a) and $\delta$-$NbNi_3$ (b) by three models: novel model (blue maker), light model (green maker), and standard model (orange maker) from SIPFENN and Alignn (red maker) at 298 K of the Nb-Ni system, in comparison with the DFT results (purple maker) and value = 0 (black line).

**Fig. S3.** shows the presently predicted $\Delta H_{form}$ of all configurations of $\mu$-$Nb_7Ni_6$ (a) and $\delta$-$NbNi_3$ (b) comparing with the desired DFT values.

**Fig. S4.** shows the presently predicted convex hull of $\Delta H_{form}$ of $\mu$-$Nb_7Ni_6$ (a) and $\delta$-$NbNi_3$ (b) comparing with the DFT results.

## Supplemental Figures:

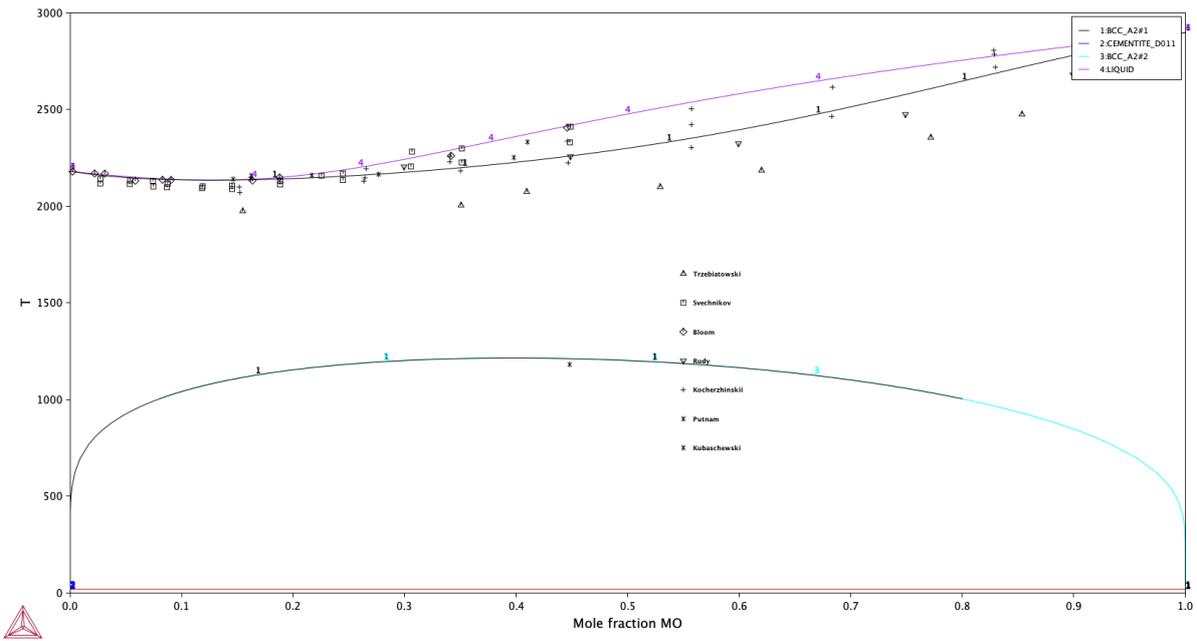

Fig. S1. Calculated Cr – Mo phase diagram by adopting the thermodynamic description by Frisk et al. [39].

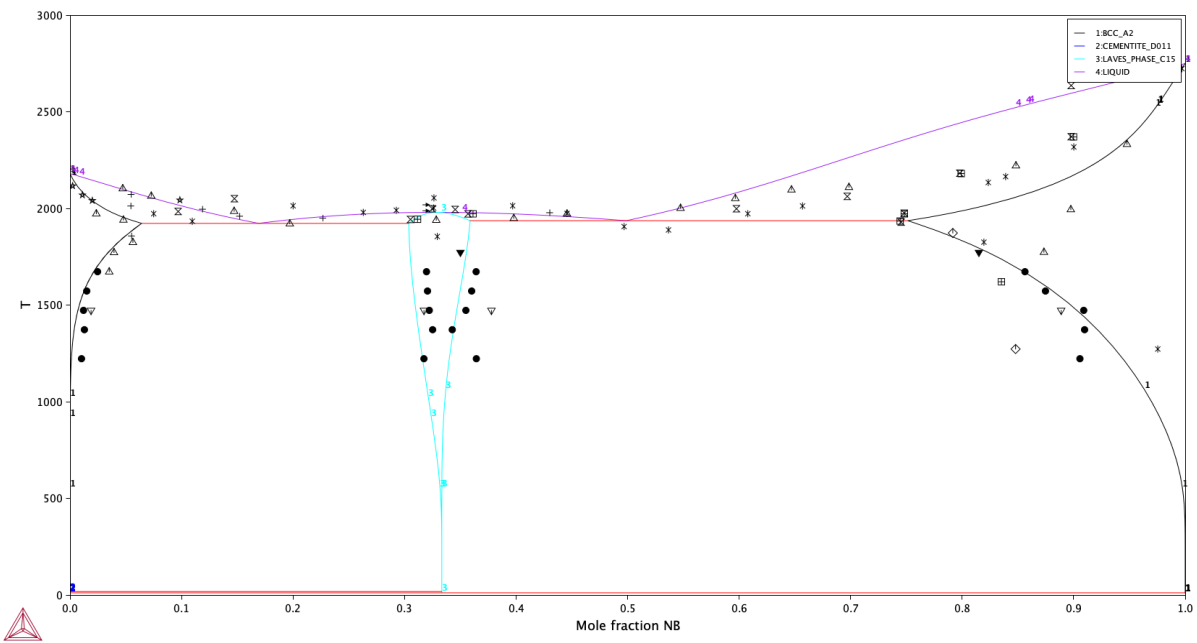

Fig. S2. Calculated Cr – Nb phase diagram by adopting the thermodynamic description by Peng et al. [16].

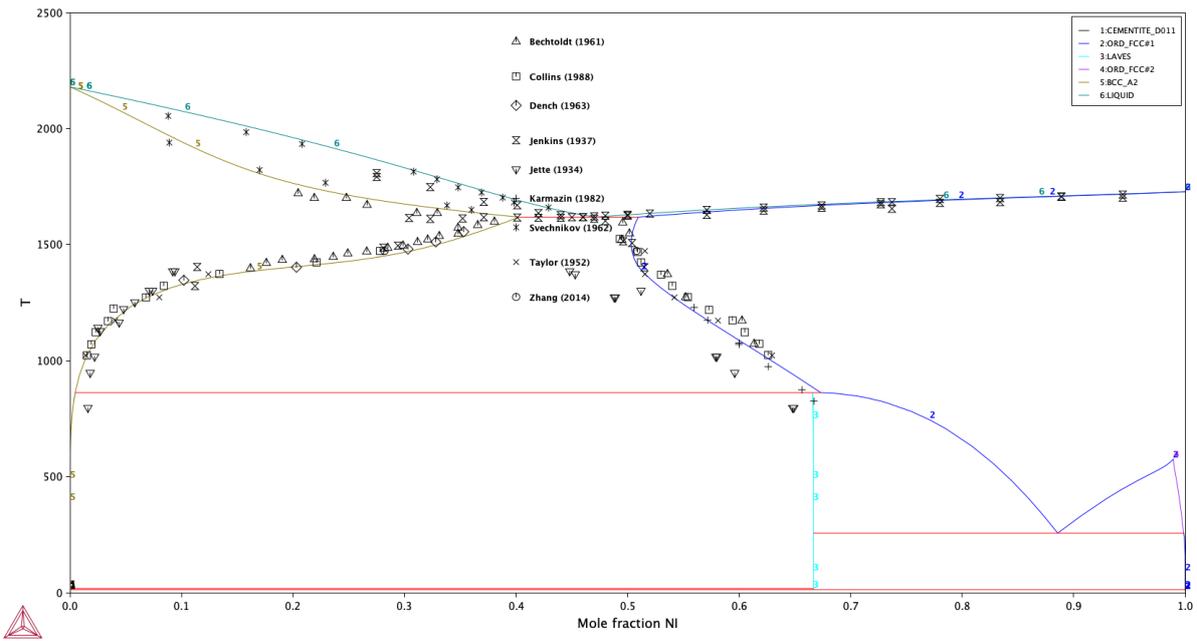

Fig. S3. Calculated Cr – Ni phase diagram by adopting the thermodynamic description by Tang et al. [56] but with remodeling of Laves_C14 using a complete three-sublattice model.

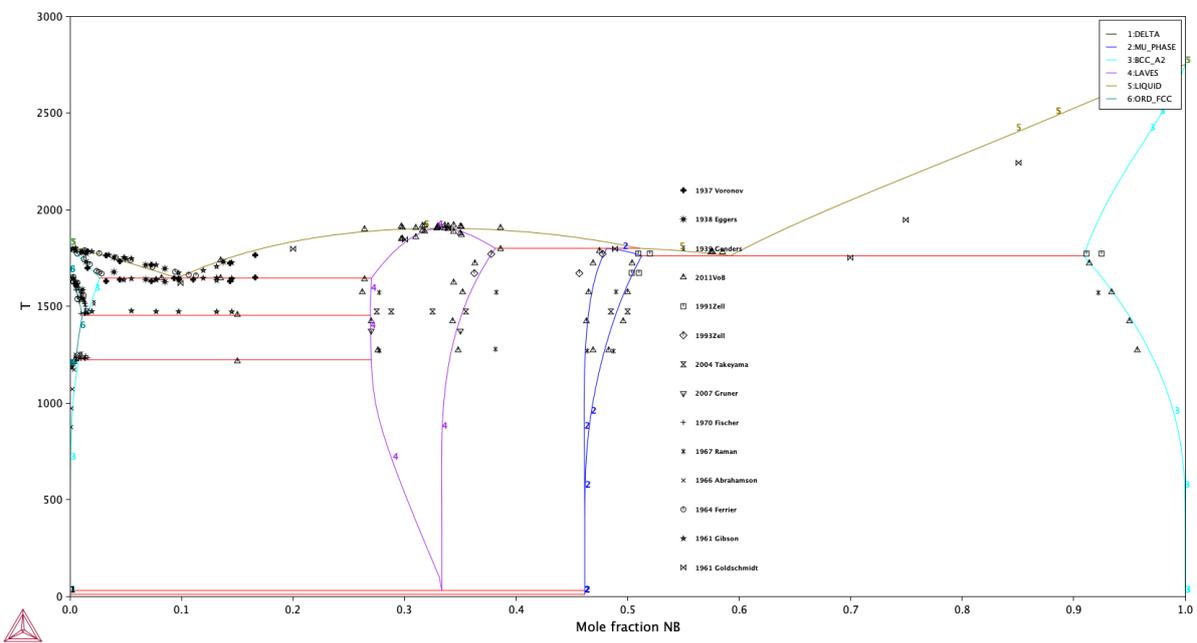

Fig. S4. Calculated Fe – Nb phase diagram by adopting the thermodynamic description by Sun et al. [72].

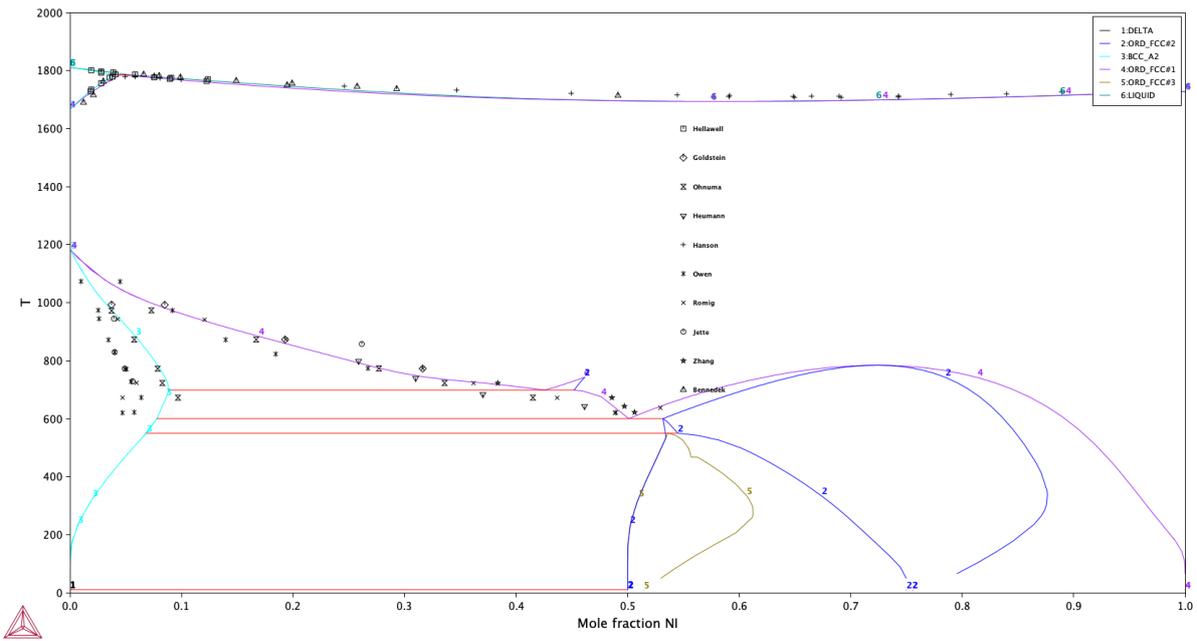

Fig. S5. Calculated Fe – Ni phase diagram by adopting the thermodynamic description by Sun et al. [72].

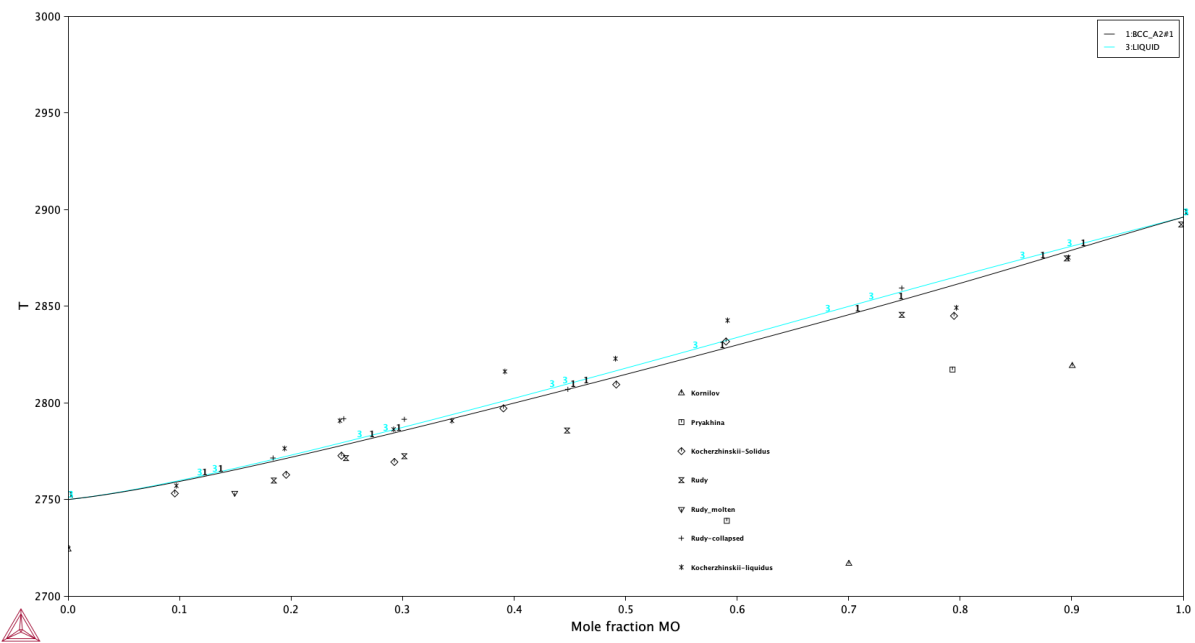

Fig. S6. Calculated Mo – Nb phase diagram by adopting the thermodynamic description by Yen et al. [83].

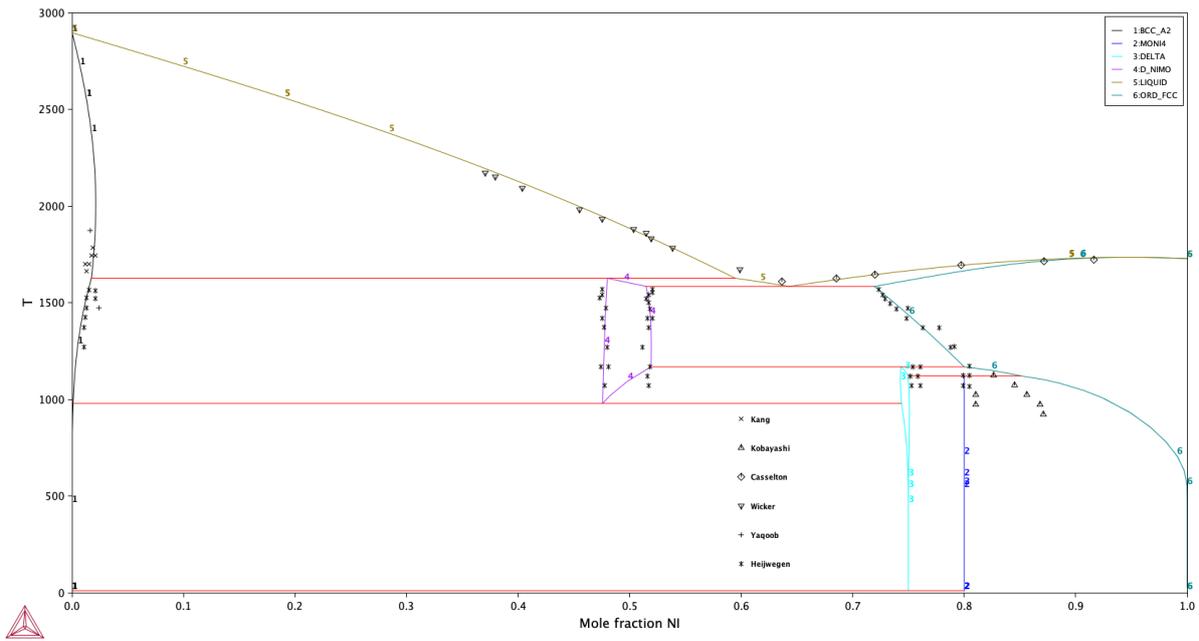

Fig. S7. Calculated Mo – Ni phase diagram by adopting the thermodynamic description by Lei et al. [17] but with changing from the two-sublattice to three-sublattice model for the δ phase.

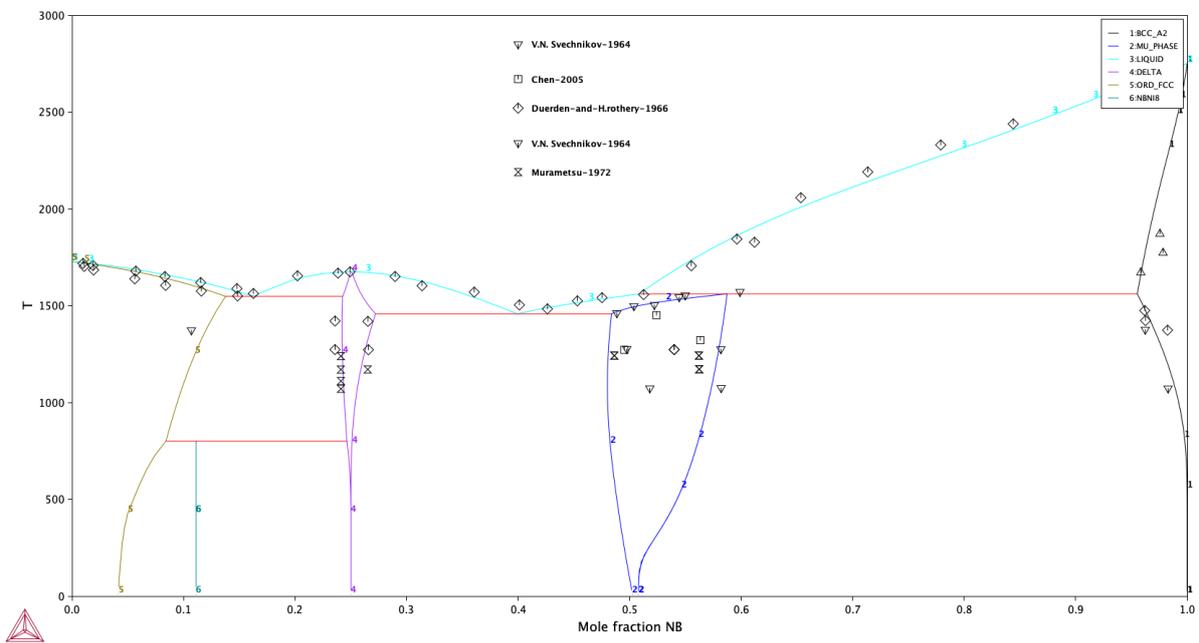

Fig. S8. Calculated Nb – Ni phase diagram by adopting the thermodynamic description by Sun et al. [93].

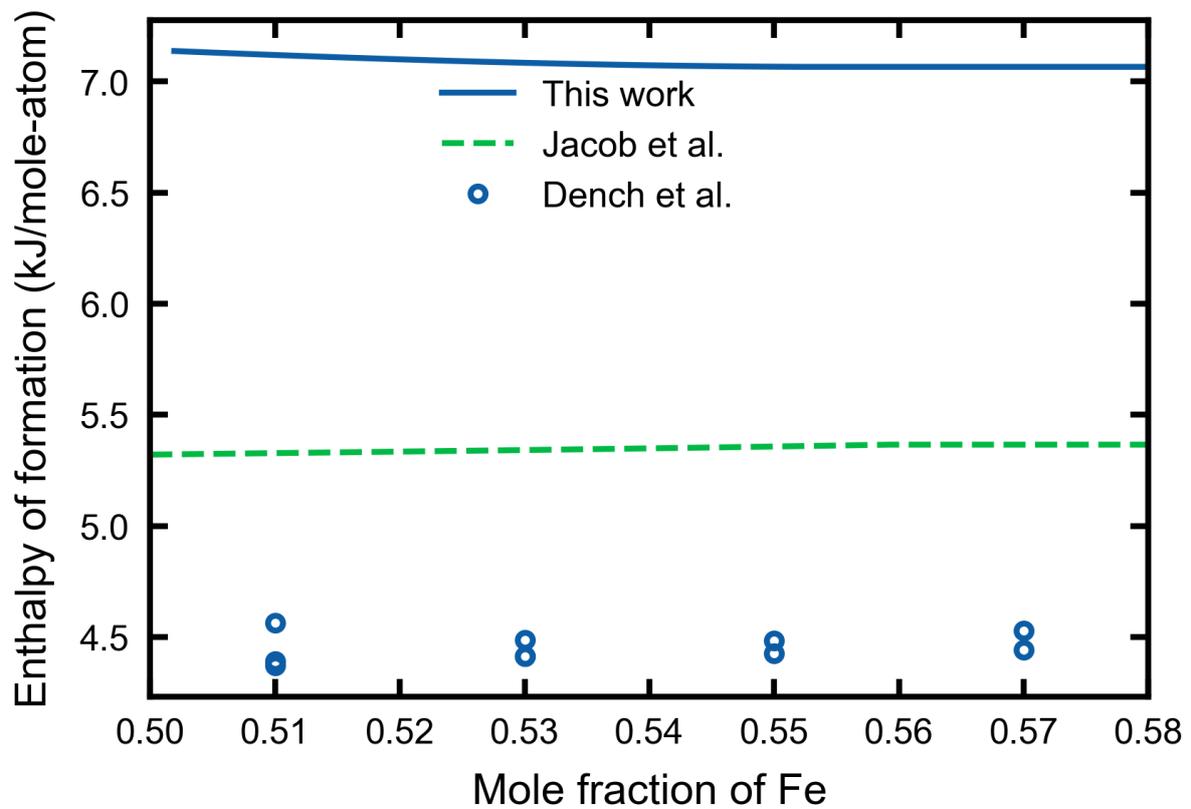

Fig. S9 present enthalpy of formation at 923 K at 50.0 to 58.0 at. % Fe calculated from both this study and the modeling from Jacob et al. [14] with experimental data from Dench et al. [102]. Both of the modeling work match well with the experimental data from Dench et al. [102]. The present work shows 0.55 kJ/mol-atom difference comparing with the data from Dench et al. [102]. The modeling from Jacob et al. [14] shows 0.89 kJ/mol-atom difference comparing with the data from Dench et al. [102].

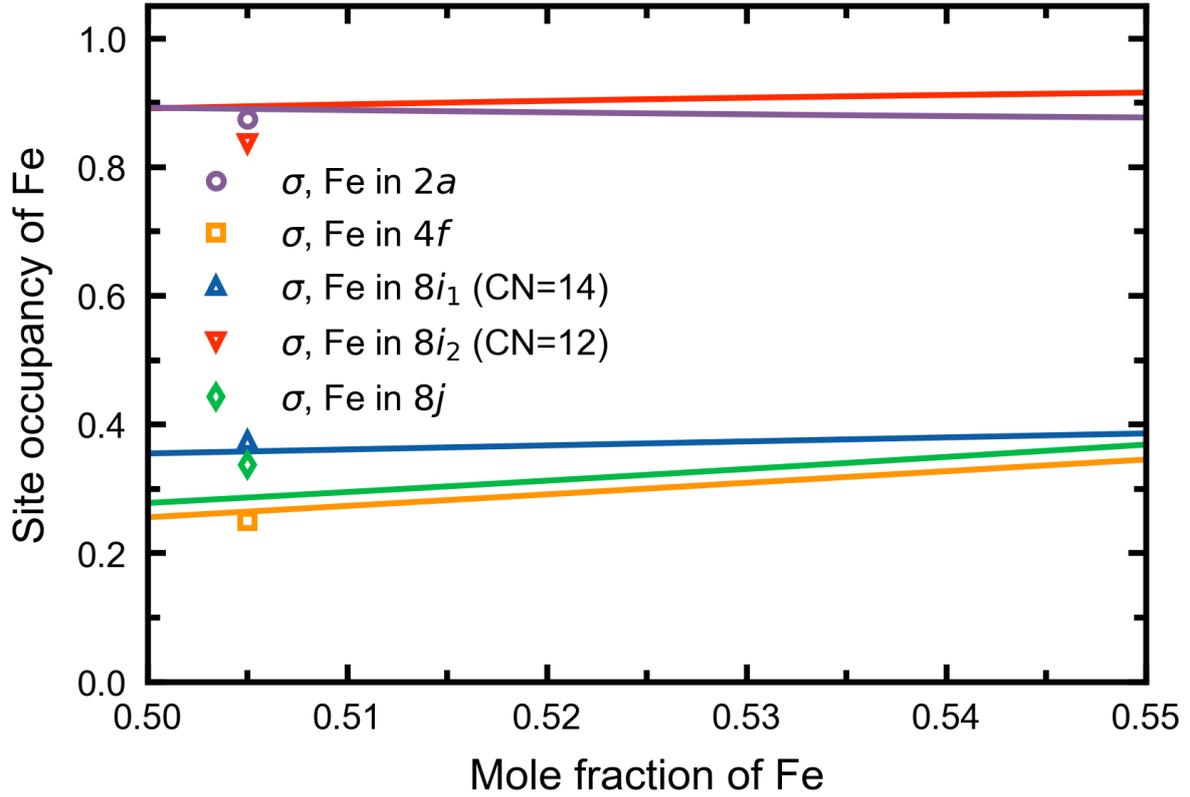

Fig. S10. Predicted $\Delta H_{form}$ of endmembers of μ-Nb$_7$Ni$_6$ (a) and δ-NbNi$_3$ (b) by three models: novel model (blue maker), light model (green maker), and standard model (orange maker) from SIPFENN and Alignn (red maker) at 298 K of the Nb-Ni system, in comparison with the DFT results (purple maker) and value = 0 (black line).

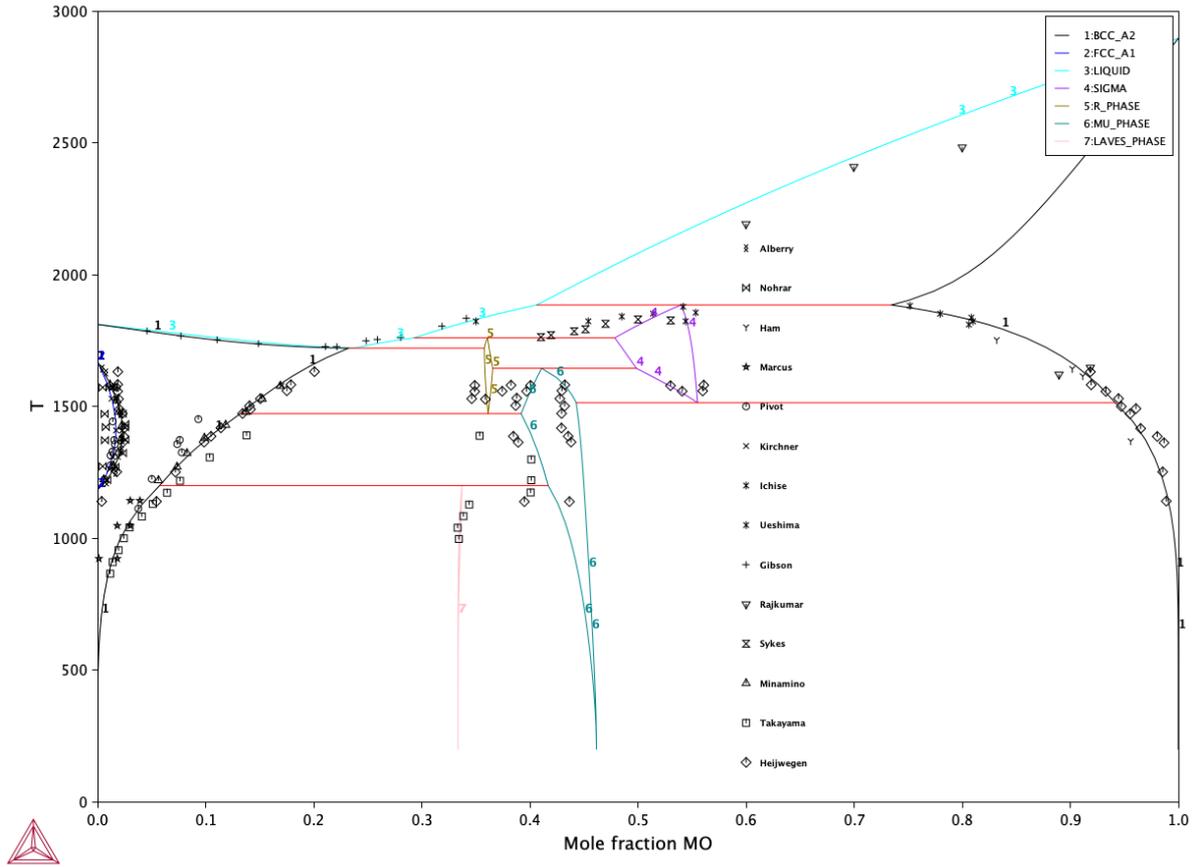

Fig. S11. Calculated Fe – Mo phase diagram by adopting the thermodynamic description in the present work.

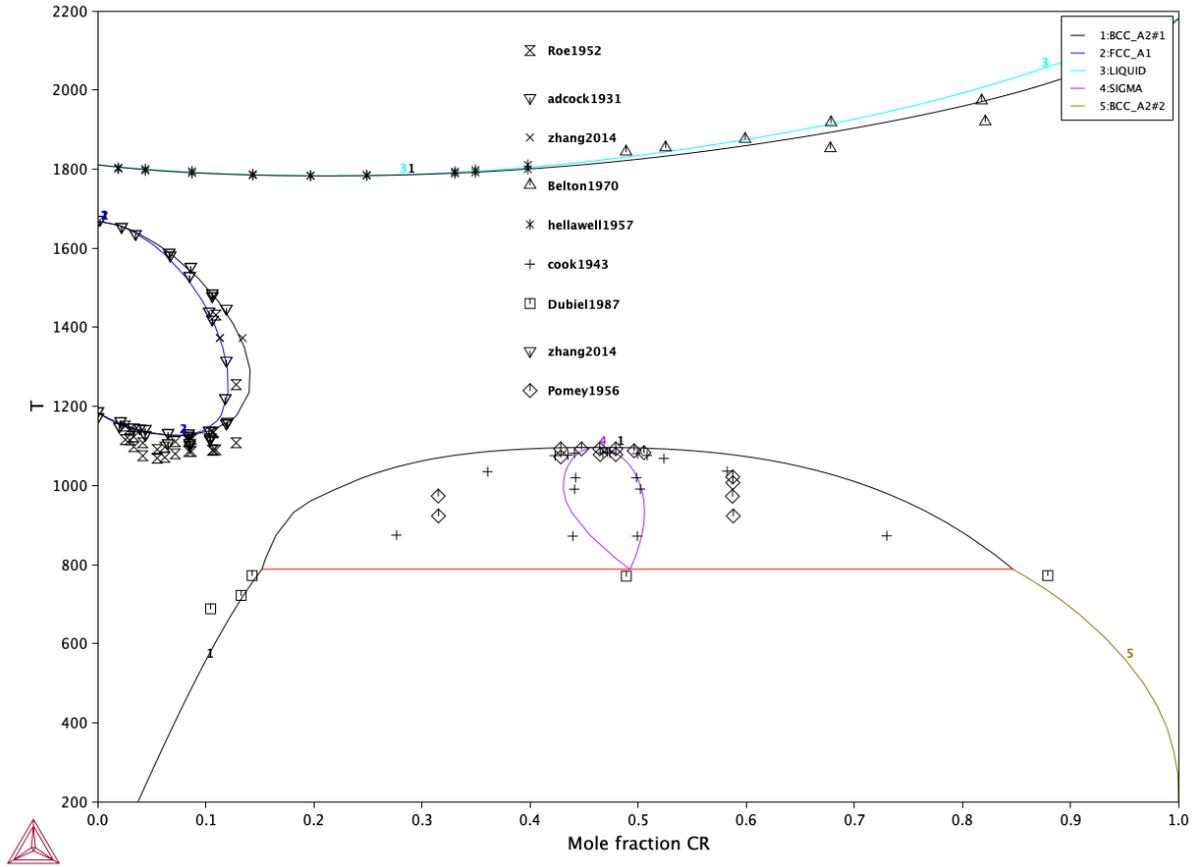

Fig. S12. Calculated Cr – Fe phase diagram by adopting the thermodynamic description in the present work.